\documentclass[aps,prr,english,twocolumn,preprintnumbers,amsmath,amssymb,nofootinbib,superscriptaddress,longbibliography,noeprint]{revtex4-2}

\pdfoutput=1

\usepackage[latin1]{inputenc}
\usepackage{graphicx}
\usepackage{color}
\usepackage{bbm}
\usepackage{float}
\usepackage{amssymb}
\usepackage{physics}
\usepackage{amsmath}
\usepackage{tabularx}
\usepackage[dvipsnames]{xcolor}
\usepackage{titlesec}
\usepackage{enumitem} % use to set margin of enumerate env

\usepackage{dsfont}

\usepackage[colorlinks,citecolor=blue]{hyperref}

\def\0#1#2{\frac{#1}{#2}}

\def\s0#1#2{\mbox{\small{$ \frac{#1}{#2} $}}}

\def\CZ{{\mathcal Z}}
\def\CD{{\mathcal D}}

\newcommand{\beq}{\begin{equation}}
\newcommand{\eeq}{\end{equation}}
\newcommand{\bea}{\begin{eqnarray}}
\newcommand{\eea}{\end{eqnarray}}

\newcommand{\tpderiv}[2]{\left(\frac{\partial #1}{\partial #2}\right)}

\titleformat{\subsubsection}    
{\bfseries\itshape\filcenter}{\thesubsubsection .}{1em}{}
\usepackage{babel}
\makeatother
\begin{document}

\title{Thermodynamics of spin-$1/2$ fermions on coarse temporal lattices \\ using automated algebra}

\author{K. J. Morrell}
\affiliation{Department of Physics and Astronomy, University of North Carolina,
  Chapel Hill, North Carolina 27599, USA}

\author{A. J. Czejdo}
\affiliation{Department of Physics and Astronomy, University of North Carolina,
  Chapel Hill, North Carolina 27599, USA}

\author{N. Carter}
\affiliation{Department of Computer Science, University of North Carolina,
	Chapel Hill, North Carolina 27599, USA}

%\author{Y. Hou}
%\affiliation{Department of Physics and Astronomy, University of North Carolina,
%	Chapel Hill, North Carolina 27599, USA}
  
\author{J. E. Drut}
\affiliation{Department of Physics and Astronomy, University of North Carolina,
  Chapel Hill, North Carolina 27599, USA}

\begin{abstract}
Recent advances in automated algebra for dilute Fermi gases in the virial expansion, where coarse 
temporal lattices were found advantageous, motivate the study of more general computational schemes
that could be applied to arbitrary densities, beyond the dilute limit where the virial expansion is
physically reasonable. We propose here such an approach by developing
what we call the Quantum Thermodynamics Computational Engine (QTCE).
In QTCE, the imaginary-time direction is discretized and the interaction is accounted for via a quantum cumulant 
expansion, where the coefficients are expressed in terms of noninteracting expectation values.
The aim of QTCE is to enable the systematic resolution of interaction effects at fixed temporal
discretization, as in lattice Monte Carlo calculations, but here in an algebraic rather than numerical fashion.
Using this approach, in combination with numerical integration techniques (both known and alternative
ones proposed here), we explore the thermodynamics of spin-$1/2$ fermions,
focusing on the unitary limit in 3 spatial dimensions, but also exploring the effects of continuously varying the
spatial dimension below 3. We find that, remarkably, extremely coarse temporal lattices,
when suitably renormalized using known results from the virial expansion, yield stable partial sums for 
QTCE's cumulant expansion which are qualitatively and quantitatively correct in wide regions, compared with
known experimental results.
\end{abstract}

\maketitle

%%%%%%%%%%%%%%%%%%%%%%%%%%%%%%%%%%%%%%%%%%%%%%%%%%%%%%%%%%%%%%%%
\section{Introduction}

Approaches to the quantum many-body problem, in particular at finite temperature, can be roughly divided into
two large classes: analytical methods (e.g. perturbation theory, mean-field approaches, etc) and numerical methods (e.g. all the 
well-known flavors of quantum Monte Carlo; see e.g.~\cite{DrutLattice, BergerSignProb}). Naturally, the former do require some numerical evaluation at the very end, while 
the latter do require some analytic development (i.e. rewriting the problem in a way amenable to modern computers) before any 
actual numerical computations are carried out.

As modern computers stay on the path set by Moore's law, interesting approaches have emerged that bridge
between the above two paradigms. This is the idea of automated algebra: the use of computers to perform
extremely lengthy derivations (of a scale that would be impossible to accomplish by hand in a human lifetime) in order to explore
the quantum many-body problem analytically as far as possible, before proceeding to numerical evaluation at the end.
The application of this concept has seen examples in nuclear physics (see e.g.~\cite{StevensonJMPC, ARTHUIS2019202, ARTHUIS2021107677, Tichai2022}) and quantum chemistry (see e.g.~\cite{Hirata2006}) for some time and more
recently in ultracold fermions (see below).
One may regard the idea of automated algebra as merely extending conventional analytic methods, which is correct in some regard 
but we have found, in the research presented here and elsewhere, that conventional numerical approaches also inform automated 
algebra methods in relevant and useful ways.

As a specific example pursued by our group, automated algebra has successfully pushed the boundaries of the quantum 
virial expansion (VE) for spin-$1/2$ Fermi gases in a wide range of settings (see e.g.~\cite{Hou2020PRL, PhysRevA.102.033319, PhysRevResearch.3.033099} and~\cite{condmat7010013} for a recent review), managing to obtain precise estimates of up to the fifth order coefficient.
Encouraged by the success of automated algebra for the VE, here we explore a different route that is not a priori restricted
to low-density regimes. We propose an approach that we call the Quantum Thermodynamics Computational Engine (QTCE)
whereby the imaginary-time direction is discretized and the interaction effects are accounted for via a quantum cumulant 
expansion. In turn, the quantum cumulants at a given order are calculated using the corresponding moments (and their
lower-order counterparts) in terms of noninteracting expectation values. In its present form, the objective of QTCE is to 
enable the calculation of fundamental thermodynamic quantities such as the pressure, density, and isothermal compressibility, 
in a systematic resolution of interaction effects at fixed temporal discretization. As in our previous VE investigations, we focus
on short-range, contact interactions, such that the matrix elements have no momentum dependence (aside from total momentum
conservation, of course), which facilitates the exploration of novel approaches to the evaluation of the relevant 
multidimensional integrals using algebraic methods.

The remainder of this paper is organized as follows. Section \ref{sec:formalism} introduces the Hamiltonian of our system with 
details of the discretization and outlines the expansion of the partition function in terms of multivariate quantum cumulants. 
Section \ref{sec:compDetails} then connects this expansion with the noninteracting generating functional of density correlation 
functions before describing the QTCE implementation. Section \ref{sec:results} presents our thermodynamic results for spin-1/2 
fermions, focusing on the universal limit of the unitary Fermi gas. In Sec.~\ref{sec:NtauEffects} we discuss two systematic effects 
related to coarse temporal lattices and the choice of integrator approach. Finally, we conclude in Sec.~\ref{sec:Conclusion&Outlook}, where we also comment on the generalizability of our method.

%%%%%%%%%%%%%%%%%%%%%%%%%%%%%%%%%%%%%%%%%
\section{Formalism}
\label{sec:formalism}
%%%%%%%%%%%%%%%%%%%%%%%%%%%%%
\subsection{Model, partition function, and discretization}
We consider a non-relativistic, spin-$1/2$ fermionic system with Hamiltonian $\hat H = \hat T + \hat V$, where 
\beq
\hat T = \sum_{s=\uparrow,\downarrow} \int d^d x \text{ }\hat \psi_s^\dagger({\bf x}) \left(- \frac{\hbar^2\nabla^2}{2m} \right) \hat \psi_s({\bf x}),
\eeq
is the kinetic energy in terms of spatial dimension $d$, particle mass $m$, and field operators $\hat \psi_s^\dagger$, $\hat \psi_s$ for spin $s$, and
\beq
\hat V = -g \int d^d x \text{ }\hat n^{}_\uparrow({\bf x})\hat n^{}_\downarrow({\bf x}),
\eeq
is a contact interaction with coupling constant $g$ and particle density operators $\hat n^{}_s=\hat \psi_s^\dagger\hat\psi^{}_s$. 
To proceed, we let $\hbar=k_B=m=1$ and regularize the interaction by discretizing spacetime using a spatial lattice spacing 
$\ell$ such that the potential energy takes the form
\beq
\hat V = -g_{lat} \sum_{\bf x} \hat n^{}_\uparrow({\bf x})\hat n^{}_\downarrow({\bf x}),
\eeq
where the operators are now dimensionless and $g_{lat} = g/\ell^d$.

To study thermodynamics, the central quantity of interest is the grand-canonical partition function of the unpolarized, interacting system,
\beq
\mathcal Z = \text{Tr} \left[ e^{-\beta (\hat H - \mu \hat N)} \right],
\eeq
where $\beta$ is the inverse temperature, $\mu$ is the chemical potential (common to both spin degrees of freedom, 
as we focus on unpolarized matter here), and $\hat N$ is the total particle number. 

The central challenging point of quantum thermodynamics is that $\hat T$ and $\hat V$ do not commute.
To address that problem, we follow a common route and introduce a temporal discretization 
$\beta = \tau N^{}_\tau$ where $N^{}_\tau$ is a positive integer and $\tau$ is the spacing, 
which allows us to separate the kinetic and potential energy exponential factors via a (symmetric) Trotter-Suzuki 
factorization, namely
\beq
e^{-\tau (\hat H - \mu \hat N)}  = e^{-\tau \hat K^{}_0/2} e^{-\tau \hat V} e^{-\tau \hat K^{}_0/2} + O(\tau^3),
\eeq
where $\hat K^{}_0 = \hat T - \mu \hat N$. The net result on the partition function is the following approximation
(valid up to $O(\tau^2)$ once all $N_\tau$ time slices are accounted for):
\beq
\mathcal Z = \text{Tr} \left[ e^{-\beta (\hat H - \mu \hat N)} \right]
\simeq \text{Tr} \left[ e^{-\tau \hat K^{}_0} e^{-\tau \hat V} e^{-\tau \hat K^{}_0} e^{-\tau \hat V} \dots \right],
\eeq
where there are $N^{}_\tau$ factors. Since $\tau= \beta/N^{}_\tau$, it follows that increasing $N^{}_\tau$ improves the accuracy of the final answer while also increasing the computational cost.

On a spatial lattice, we may use the property $\hat n^2 = \hat n$, valid for dimensionless, lattice fermion density operators, to write
\bea
\label{Eq:ExptauV}
e^{-\tau \hat V} &=& \prod_{\bf x} \left[1 + C \hat n^{}_\uparrow({\bf x})\hat n^{}_\downarrow({\bf x})\right] \\
&=& \sum_{k}\frac{\lambda^k}{k!}\hat V_k,
\eea
for $k\geq0$ where we have introduced an arbitrary parameter $\lambda$ (see below) and
\beq
\label{Eq:Vsubk}
\hat V_k = C^k \sum_{{\bf x}_1 \neq {\bf x}_2 \neq \dots \neq {\bf x}_k}
\hat n^{}_2 ({\bf x}_1)\hat n^{}_2 ({\bf x}_2) \dots \hat n^{}_2 ({\bf x}_k),
\eeq
with $C = (e^{\tau g} - 1)/\lambda$, $\hat n^{}_2({\bf x}) = \hat n^{}_\uparrow({\bf x})\hat n^{}_\downarrow({\bf x})$, and 
$\hat V_0 = \openone$.
When envisioning taking the large-$N^{}_\tau$ limit at constant $\beta$, it is useful to set $\lambda = 1/N^{}_\tau$,
such that, in that limit, one has $C \to \beta g$, i.e. $C$ approaches a well-defined limit in terms of the parameters 
of the theory. In practice, $C$ is set by renormalization (more on this below), such that the choice of $\lambda$
should be immaterial (up to, possibly, numerical artifacts). We will also see below that choosing $\lambda = 1/N^{}_\tau$
makes sense from the point of view of the scaling of the number of terms in the cumulant expansion presented below 
as $N^{}_\tau$ is increased.

%%%%%%%%%%%%%%%%%%%%%%%%%%%%%
\subsection{Moment and cumulant expansions}

We make use of the interaction expansion of Eq.~(\ref{Eq:ExptauV}) by expanding the partition function into 
a quantum moment expansion:
\beq
\label{Eq:Wexpansion}
\frac{\mathcal Z(\lambda)}{\mathcal Z_0} = 
\sum_{\{n\}=0}^{\infty}
\frac{\lambda^{n^{}_1 + n^{}_2 +\dots + n^{}_{N^{}_\tau}}}{n^{}_1! n^{}_2! \dots n^{}_{N^{}_\tau}!} W_{n^{}_1,n^{}_2,\dots,n^{}_{N^{}_\tau}},
\eeq
where $\mathcal Z_0$ is the partition function of the non-interacting system and 
\bea
\label{Eq:Wdef}
W_{n^{}_1,n^{}_2,\dots,n^{}_{N^{}_\tau}} &=& 
\frac{1}{\mathcal Z_0}
\text{Tr} 
\left[ 
e^{-\tau \hat K^{}_0} \hat V_{n^{}_1} 
e^{-\tau \hat K^{}_0} \hat V_{n^{}_2}
\dots \right],\nonumber \\
\eea
are the quantum moments. Based on the above moment expansion, the following cumulant expansion results, 
which is one of the cornerstones of the QTCE, as described below:
\beq
\label{Eq:Kexpansion}
\ln \left(\frac{\mathcal Z(\lambda)}{\mathcal Z_0}\right) = 
\sum_{\{n\}=0}^{\infty} 
\frac{\lambda^{n^{}_1 + n^{}_2 +\dots + n^{}_{N^{}_\tau}}}{n^{}_1! n^{}_2! \dots n^{}_{N^{}_\tau}!}
\kappa_{n^{}_1,n^{}_2,\dots,n^{}_{N^{}_\tau}},
\eeq
where $\kappa_{0,0,\dots,0} = 0$ for any value of $N^{}_\tau$, because $W_{0,0,\dots,0} = 1$ for all $N^{}_\tau$. It is this expansion that gives us access to thermodynamic observables via the fundamental relation
\beq
\label{Eq:P-lnZ}
\beta(P-P_0)V=\ln \left(\frac{\mathcal Z}{\mathcal Z_0}\right),
\eeq
where ($P_0$) $P$ is the pressure of the (non-)interacting system and $V$ is the $d$-dimensional volume.

The series form of Eq.~(\ref{Eq:Kexpansion}) naturally allows one to study the effect of including successively higher-order 
terms, when the series is truncated at some order $N^{}_\kappa$. Thus, our method has two parameters, namely 
$N^{}_\tau$ and $N^{}_\kappa$, which can be systematically increased. In this work, we favor increasing $N^{}_\kappa$ while keeping $N^{}_\tau$ relatively small (thus the 
``coarse temporal lattices" advertised in the title). Note that taking these limits in reverse order amounts to a perturbative
approach, while increasing $N^{}_\kappa$ first, at constant $N^{}_\tau$ (as we propose), is formally closer to lattice Monte Carlo 
calculations.

Using
\beq
\beta P_0 V = \left(\frac{L}{\lambda_T}\right)^d I_0(z), 
\eeq
where $\lambda_T = \sqrt{2\pi\beta}$ is the thermal wavelength, and
\beq
I_0(z) = \frac{1}{(2\pi)^{d/2}} \int dx \ln (1 + z e^{-\frac{x^2}{2}}),
\eeq
we can reformulate Eq.~(\ref{Eq:Kexpansion}) as
\bea
\frac{P}{P_0} &=& 1 + \sum_{\{n\}=1}^{\infty} 
\lambda^{n^{}_1+n^{}_2 + \dots + n^{}_{N^{}_\tau}}
\frac{\bar \kappa_{n^{}_1,n^{}_2,\dots,n^{}_{N^{}_\tau}}}{n^{}_1!n^{}_2!\dots n^{}_{N^{}_\tau}!}, \label{eq:Kbar-expansion} \\
&=& 1 + \lambda \bar K^{}_1 + \lambda^2 \bar K^{}_2 +\dots, \label{eq:bigKbar}
\eea
where
\beq
\bar \kappa_{n^{}_1,n^{}_2,\dots,n^{}_{N^{}_\tau}} = \left(\frac{\lambda_T}{L}\right)^d \frac{\kappa_{n^{}_1,n^{}_2,\dots,n^{}_{N^{}_\tau}}}{I_0(z)},
\eeq
and we define the ``assembled cumulants" as
\bea
\bar K^{}_1 &=& \bar \kappa_{1,0,\dots,0} + \bar \kappa_{0,1,0,\dots,0} + \dots + \bar \kappa_{0,0,\dots,1}, \\
\bar K^{}_2 &=& \frac{1}{2}\bar \kappa_{2,0,\dots,0} + \frac{1}{2}\bar \kappa_{0,2,0,\dots,0} + \dots  \nonumber \\
&& + \frac{1}{2}\bar \kappa_{0,0,\dots,2} + \bar \kappa_{1,1,0,\dots,0} + \dots,
\eea
and so on. 

Note that since $\bar \kappa_{1,0,\dots,0} = \bar \kappa_{0,1,0,\dots,0} = \dots = \bar \kappa_{0,0,\dots,1}$, we can write $\bar K^{}_1 = N^{}_\tau \bar \kappa_{1,0,\dots,0}$, which explicitly shows that $\bar K^{}_1$ is linear in $N^{}_\tau$, and this
dependence is then cancelled in the expansion by the factor $\lambda = 1/N^{}_\tau$ in Eq.~(\ref{eq:bigKbar}). 
Similarly, it is easy to see that there are $\propto N^{2}_\tau$ terms in $\bar K^{}_2$, whose number is balanced by 
the $\lambda^2 = 1/N^{2}_\tau$ dependence in Eq.~(\ref{eq:bigKbar}). The same reasoning applies to all orders in the 
expansion of Eq.~(\ref{eq:bigKbar}).

These $\bar K^{}_n$ quantities are useful for analyzing the system and are the main output of QTCE, which returns results up to a given cumulant order $N^{}_\kappa$, corresponding to the upper limit on the summation in Eq.~(\ref{eq:Kbar-expansion}). We 
expect the $\kappa_{n^{}_1,n^{}_2,\dots,n^{}_{N^{}_\tau}}$ to be extensive quantities proportional to $\left({L}/{\lambda_T}\right)^d$ and 
therefore all $\bar \kappa_{n^{}_1,n^{}_2,\dots,n^{}_{N^{}_\tau}}$ to be intensive.

The cumulants can generally be expressed in terms of the moments via recursive formulas, such that to calculate a cumulant at a 
given order, one needs all of the moments up to and including that order \cite{RecursiveCumulants}. For example, for $N^{}_\tau=1$, 
\beq
\label{Eq:CumulantRecursive}
\kappa_n = W_n - \sum_{m=1}^{n-1} C^{n-1}_{m-1} \kappa_m W_{(n-m)},
\eeq
where $C^{n}_k= n!/(k!(n-k)!)$ is the binomial coefficient. 

Since Eq.~(\ref{Eq:P-lnZ}) establishes that $\ln (\mathcal Z/\mathcal Z_0)$ is proportional to the volume $V$, 
and $V$ and $\lambda$ are independent variables, it must be that within each cumulant $\kappa_{n}$ 
the only terms that contribute are those with the proper spatial volume scaling $\propto V$. Since this is true 
for all $n$, we see that in 
Eq.~(\ref{Eq:CumulantRecursive}) the role of the terms in the sum is to cancel out the unnecessary terms 
in $W_{n}$. In practice, we therefore proceed simply by calculating $W_{n}$ and retaining only the terms 
with linear $V$ scaling. In terms of Feynman diagrams, this amounts to keeping only the connected diagrams, 
as indicated by the linked-cluster theorem (see e.g. Ref.~\cite{Negele1998Quantum}). Thus, we are able to eliminate a large 
number of terms early on in the calculation, thereby reducing the computational cost.

%%%%%%%%%%%%%%%%%%%%%%%%%%%%%
\section{Computational details}
\label{sec:compDetails}
\subsection{Toward calculating the moments $W_{n^{}_1,n^{}_2,\dots,n^{}_{N^{}_\tau}}$ with generating functionals}

In Eq.~(\ref{Eq:Wdef}) we saw that in order to calculate the moment $W_{n^{}_1,n^{}_2,\dots,n^{}_{N^{}_\tau}}$ we need
\bea
&\text{Tr} 
\left[ 
e^{-\tau \hat K^{}_0} \hat V_{n^{}_1} 
e^{-\tau \hat K^{}_0} \hat V_{n^{}_2}
\dots \right] = \nonumber \\ 
&\int \mathcal D x \;
\text{Tr} 
\left[ 
e^{-\tau \hat K^{}_0} \prod_{k=1}^{n^{}_1} \hat n^{}_2 ({\bf x}^{}_k)
e^{-\tau \hat K^{}_0} \prod_{k=1}^{n^{}_2} \hat n^{}_2 ({\bf y}^{}_k)
\dots \right],\nonumber \\
\eea
where we define
\beq
\label{Eq:XIntegral}
\int \mathcal D x \equiv \sum_{{\bf x}^{}_1 \neq \dots \neq {\bf x}^{}_{n^{}_1}} C^{n^{}_1} 
\sum_{{\bf y}^{}_1 \neq \dots \neq {\bf y}^{}_{n^{}_2}} C^{n^{}_2}\dots,
\eeq
to represent the spatial integration that is part of the simplification and integration modules of QTCE.

Next, we turn to calculating the noninteracting expectation value 
\beq
\label{Eq:NIExpVal}
\frac{1}{\mathcal Z_0}
\text{Tr} 
\left[ 
e^{-\tau \hat K^{}_0} \prod_{k=1}^{n^{}_1} \hat n^{}_2 ({\bf x}^{}_k)
e^{-\tau \hat K^{}_0} \prod_{k=1}^{n^{}_2} \hat n^{}_2 ({\bf y^{}}_k)
\dots \right],
\eeq
using a noninteracting generating functional and automated algebra.

%%%%%%%%%%%%%%%%%%%%%%%%%%%%%
\subsection{Using the non-interacting generating functional of density correlation functions}

The noninteracting generating functional of density correlation functions is defined as
\beq
\mathcal Z_0[j] = \text{Tr} \left[ e^{-\tau \hat K^{}_0} e^{\hat V_{\text{ext},1}} e^{-\tau \hat K^{}_0} e^{ \hat V_{\text{ext},2}}\dots e^{-\tau \hat K^{}_0} e^{ \hat V_{\text{ext},N^{}_\tau}} \right],
\eeq
where we couple our source $j({\bf x},t)$ to the density at each spacetime point via
\beq
\hat V_{\text{ext},t} = \sum_{\bf x} \, j({\bf x},t) \hat n({\bf x}),
\eeq
which includes an imaginary-time coordinate $t$ to identify the various insertions of
$e^{\hat V_\text{ext}}$ in $\mathcal Z_0[j]$. Furthermore, we have assumed that our system
is placed on a lattice as a way to regularize a future interaction, such that all the quantities below, unless otherwise stated, are lattice quantities.

By taking functional derivatives of $\mathcal Z_0[j]$ with respect to $j$ at a given point
in space and time, and then dividing by $\mathcal Z_0$, we can generate noninteracting expectation values of any combination of density operators. For example,
\bea
\left\langle \prod_{\bf x} \hat n({\bf x}) \right\rangle_0 
&=& \frac{1}{\mathcal Z_0} \text{Tr} \left[ e^{-\beta \hat K^{}_0} \hat n ({\bf x}_1) \dots \hat n ({\bf x}_k)\right] \nonumber \\
&=& \frac{1}{\mathcal Z_0}\left . \frac{\delta^k \mathcal Z_0[j] }{\delta j({\bf x}_1,1)\dots\delta j({\bf x}_k,1)}  \right|_{j = 0},
\eea
where the $1$ in $({\bf x},1)$ indicates the first time slice. By taking derivatives at different time slices, the above expression can be generalized to generate expectation values of the form in Eq.~(\ref{Eq:NIExpVal}).

Because all the operators involved are now exponentials of one-body operators, the following essential simplification holds:
\bea
\label{Eq:URep}
\mathcal Z_0[j] &=& \det(1 + z U[j]) \nonumber \\
 &=& \exp\tr\ln(1 + z U[j]),
\eea
where $z = e^{\beta \mu}$ is the fugacity and $U = U_1 U_2 U_3 \dots U_{N^{}_\tau}$ is the product of the one-particle representation of the transfer matrices
\bea
[U_t]^{}_{\bf pp'} &=& \left \langle {\bf p} \right | e^{-\tau \hat T} e^{\hat V_{\text{ext},t}}  \left| {\bf p'} \right \rangle, \nonumber \\
&=& e^{-\tau p^2/(2m)} \left \langle {\bf p} \right | e^{\hat V_{\text{ext},t}}  \left| {\bf p'} \right \rangle,
\eea
where $| {\bf p} \rangle$ are single-particle states and $t$ corresponds to the time slice.

It is not difficult to show (in fact, this is a standard result in many-body physics; see e.g. 
Refs.~\cite{Negele1998Quantum, DrutLattice}) that there is another useful representation 
for the generating functional, namely
\bea
\label{Eq:MRep}
\mathcal Z_0[j] = \det M[j],
\eea
where
\bea
M[j] = 
\begin{pmatrix}
1 & 0 & \cdots & U_{N^{}_\tau} \\
-U_1 & 1 & \cdots & 0 \\
0 & -U_2 & \ddots & 0 \\
\vdots & \vdots & \ddots & \vdots \\
0 & \cdots & -U_{N^{}_\tau - 1} & 1
\end{pmatrix} .
\eea
We will refer to Eqs.~(\ref{Eq:URep}) and~(\ref{Eq:MRep}) as the $U$ and $M$ representations,
respectively, but they are often also referred to as the spatial and spacetime representations.

To simplify our notation below, we define
\beq
\label{Eq:propagator}
G(z) \equiv \frac{z e^{-\beta T}}{1 + z e^{-\beta T}},
\eeq
which is a matrix as $T$ is the single-particle matrix representation of $\hat T$, i.e.
\beq
[T]^{}_{\bf pp'} = \left \langle {\bf p} \right | \hat T \left| {\bf p'} \right \rangle = 
\delta^{}_{\bf pp'} \frac{p^2}{2m}.
\eeq
We also define
\beq
\mathcal U_{x_1,x_2,\dots,x_n} \equiv 
e^{+\beta T} \frac{\delta^n U}{\delta j(x_1) \delta j(x_2) \dots \delta j(x_n)},
\eeq
which is also a matrix, where $x_k = ({\bf x}_k, t_k)$, and similarly
\beq
\mathcal M_{x_1,x_2,\dots,x_n} \equiv 
\frac{\delta^n M}{\delta j(x_1) \delta j(x_2) \dots \delta j(x_n)}.
\eeq
It is very easy to see from the above form of $M$ that any derivative of $M$ beyond the first one
vanishes when evaluated at unequal times, i.e.
\beq
\mathcal M_{x_1,x_2} \equiv 0,
\eeq
for $t_1 \neq t_2$ and arbitrary $j$.

Similarly, it is not difficult to see that, for $t_1 = t_2$,
\beq
\mathcal U_{x_1,x_2} \equiv 0,
\eeq
for ${\bf x}_1 \neq {\bf x}_2$ and arbitrary $j$, because both $\hat V_{\text{ext},t}$ and its exponential are
diagonal matrices in coordinate space.

As usual when working with generating functionals, we set $j = 0$ at the end of the calculation. 
With these definitions one finds, for example,
\beq
\langle \hat n(x_1) \rangle_0 = 
\tr\left[
G\, \mathcal U_{x_1}
\right]
=
\tr\left[
M^{-1}\,\mathcal{M}_{x_1}
\right]
,
\eeq
and
\bea
\label{Eq:TraceExpressionsM}
\langle \hat n(x_1) \hat n(x_2) \rangle_0  &=& 
\tr \left[M^{-1}\, \mathcal M_{x_1}\right ]\tr \left[M^{-1}\, \mathcal M_{x_2}\right ]\nonumber \\ 
&-& \tr \left[M^{-1}\, \mathcal M_{x_1}\, M^{-1}\, \mathcal M_{x_2}\right ] \nonumber \\ 
&+& \tr \left[M^{-1}\,\mathcal M_{x_1,x_2}\right ],
\eea
where the last term vanishes if $t_1 \neq t_2$.
In the $U$ representation, the above becomes
\bea
\label{Eq:TraceExpressionsU}
\langle \hat n(x_1) \hat n(x_2) \rangle_0  &=& 
\tr \left[G\, \mathcal U_{x_1}\right ]\tr \left[G\, \mathcal U_{x_2}\right ]\nonumber \\ 
&-& \tr \left[G\, \mathcal U_{x_1}\, G\, \mathcal U_{x_2}\right ]  \nonumber \\ 
&+& \tr \left[G\,\mathcal U_{x_1,x_2}\right ],
\eea
and so forth, where the traces are computed in momentum space for the homogeneous system.
[In the above expressions, the time dependence appearing in the operators $\hat n$ simply indicates
the time slice where they are to be inserted, in expressions such as Eq.~(\ref{Eq:NIExpVal}).]

The case of equal times $t_1 = t_2$ requires some care for general interactions, as it may generate terms that 
diverge in the continuum limit [namely the third term in the right-hand side of Eqs.~(\ref{Eq:TraceExpressionsM}) 
and~(\ref{Eq:TraceExpressionsU})]. For a contact interaction, as is our interest here, those terms are simply
not present, by virtue of Eq.~(\ref{Eq:Vsubk}), i.e. the required equal-time correlations are only
needed at unequal spatial separation.

Thus, we see that to calculate the moments (and thus access the cumulants), one can begin by generating trace expressions for density correlations that simply involve the propagator of Eq.~(\ref{Eq:propagator}) and the first derivative of $U$. This is one of the first steps of the automated algebra code implemented by the QTCE. Note that at this stage, we have focused on
two-body local interactions (i.e. of the density-density form); generalizations of the method are discussed in Section~\ref{sec:Conclusion&Outlook}.

%%%%%%%%%%%%%%%%%%%%%%%%%%%%%
\subsection{The Quantum Thermodynamics Computational Engine}
The QTCE uses the above formalism and computational approach to evaluate the quantum cumulants of Eq.~(\ref{Eq:Kexpansion}) following the steps outlined below.
\begin{enumerate}[leftmargin=1.3em]

	\item {\sl Expectation value identification:} Determine which expectation values need to be calculated. These correspond to Eq.~(\ref{Eq:NIExpVal}) but at this stage are simply identified by the set of indices $\{n^{}_i\}$.
	
	\item {\sl Term generation:} Generate expressions for the required non-interacting expectation values using functional derivatives. At the end of this step, we have expressions as in Eq.~(\ref{Eq:TraceExpressionsU}).
	
	\item {\sl Simplification:} Carry out coordinate sums of Eq.~(\ref{Eq:XIntegral}) as well as other simplifications before assembling the cumulants. The outputs of this step are multidimensional momentum integrals, each corresponding to a diagram. They have the general form 
	\beq
	\int \CD {\bf P} \prod_{k=1}^N G[{\bf P},{\bf Q}_k,z],
	\eeq
	where (for $N^{}_\tau = 1)$
	\beq 
	G[{\bf P},{\bf Q}_k,z] = \frac{z e^{-\frac{1}{2} |{\bf Q}_k^T\cdot {\bf P}|^2}}{1+z e^{-\frac{1}{2} |{\bf Q}_k^T\cdot {\bf P}|^2}},
	\eeq
	${\bf P}^T = ({\bf p}_1,{\bf p}_2,...,{\bf p}_K)$ is a vector made out of $d$-dimensional vectors, ${\bf Q}_k$ is a $K$-dimensional vector, and
	\beq
	\CD {\bf P} \equiv d{\bf p}_1 d{\bf p}_2 ... d{\bf p}_K.
	\eeq
	
	\item {\sl Evaluation:} Evaluate the diagrams, i.e. the multidimensional integrals obtained in the previous step. This can be done with a variety of methods including a pure numerical evaluation, such as the VEGAS algorithm~\cite{VegasPaper}, or a semi-analytic approach, such as an expansion in powers of $z$, i.e. the virial expansion. Another semi-analytic approach which we have found to be useful involves an expansion in powers of $\frac{z}{1+z}$ (which is always less than 1 for all $z > 0$), as further discussed below.
	
	\item {\sl Assembly:} Once the diagrams are evaluated, their contribution to each assembled cumulant $\bar K^{}_n$ (and therefore to the pressure) are finally assembled to obtain physical results.
	
\end{enumerate}

%%%%%%%%%%%%%%%%%%%%%%%%%%%%%
\subsection{Integral evaluation by large-fugacity expansion}

When considering integrals of products of $G(z)$ functions, the main object of interest is the function
\beq
\label{Eq:Fxz}
F(x,z) = \frac{z e^{-\frac{1}{2} x^2}}{1 + z e^{-\frac{1}{2}x^2}},
\eeq
where $x^2$ represents the dispersion relation of noninteracting, nonrelativistic matter.

In the approach we advocate here, we aim to generate Gaussian integrals only. To that end, one may consider expanding 
$F(x,z)$ in powers of $z$, which effectively yields the virial expansion:
\beq
\label{Eq:FxzVE}
F(x,z) = z e^{-\frac{1}{2} x^2} \sum_{k=0}^{\infty} \left( -z\right)^k e^{-\frac{k}{2}x^2}.
\eeq
As the $e^{-\frac{k}{2}x^2}$ factor is always positive and less than or equal to $1$, this expansion is
restricted to $z < 1$, as expected and is well-known. Since our interest is in arbitrary $z$, in particular larger than $1$, we put aside this expansion and proceed in a different direction. 

\begin{figure}[h]
	\centering
	\includegraphics[width=\linewidth]{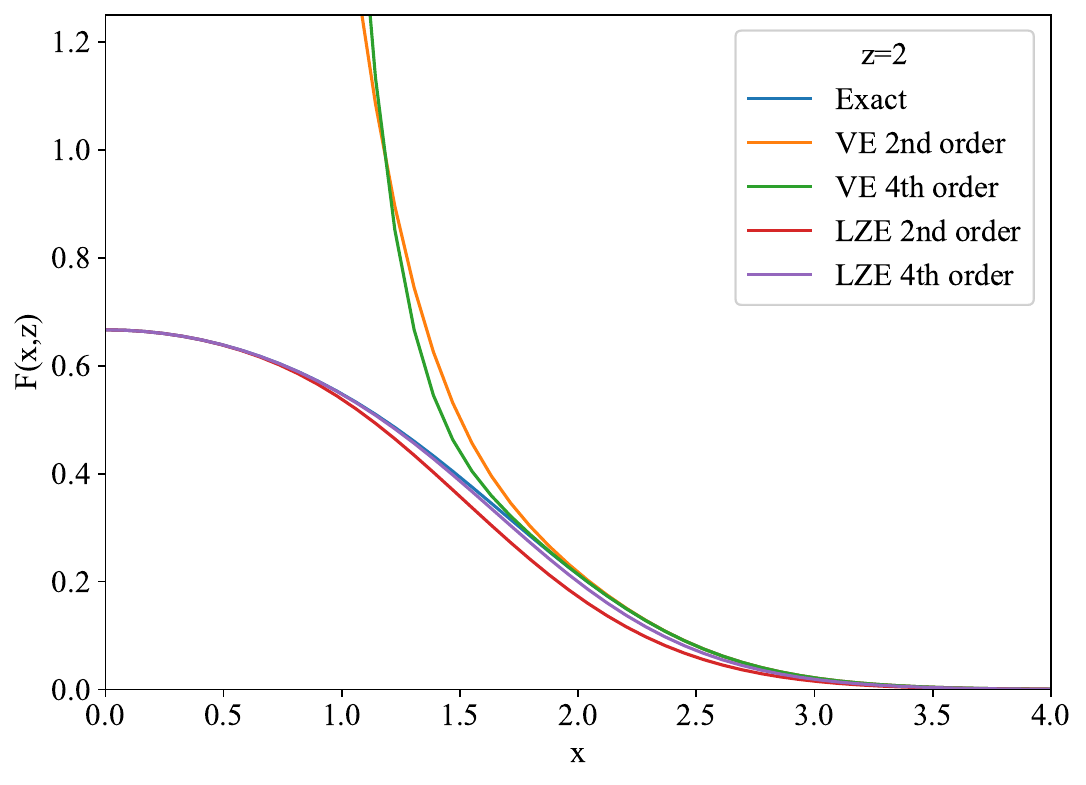}    
	\caption{
		\label{fig:LZEvsVE}
		Comparison of VE [Eq.~(\ref{Eq:FxzVE})] and LZE [Eq.~(\ref{Eq:FxzLZE})] representations of $F(x,z)$ in Eq.~(\ref{Eq:Fxz}) at second and fourth orders,
		as a function of $x$ at $z=2$. The purple line showing the fourth-order LZE is barely distinguishable, on this scale, from 
		the exact expression shown in blue.
		}
\end{figure}

To capture the general $z$ behavior, we note that regardless of the value of $z$, at large $x$ the leading term is
\beq
F(x,z) \to z e^{-\frac{1}{2} x^2},
\eeq
while at small $x$,
\beq
F(x,z) \to \frac{z}{1+z}.
\eeq
To capture these limits, we add and subtract $z$ in the denominator to obtain 
\bea
F(x,z) &=& \frac{z e^{-\frac{1}{2} x^2}}{1 + z + z (e^{-\frac{1}{2}x^2}-1)}  \\
&=&  \left(\frac{z}{1 + z}\right) \frac{e^{-\frac{1}{2} x^2}}{1 - \frac{z}{1+z}\left (1-e^{-\frac{1}{2}x^2}\right)},
\eea
which we then expand as a series in powers of $z/(1+z)$, such that
\bea
\label{Eq:FxzLZE}
F(x,z) &=& \frac{z e^{-\frac{1}{2} x^2}}{1+z} \sum_{k=0}^{\infty} \left(\frac{z}{1+z} \right)^{k} \left(1-e^{-\frac{1}{2}x^2}\right)^k.
\eea
As desired, this is a collection of Gaussians, as the virial expansion proposed earlier,
but in a different form. Moreover, the above expression is within the 
radius of convergence of the geometric series for all values of $z$ and $x$ of interest.
Since the binomial theorem dictates that 
\beq
\left(1-e^{-\frac{1}{2}x^2}\right)^k = \sum_{j=0}^{k} (-1)^j \binom{k}{j}  e^{-\frac{j}{2}x^2},
\eeq
we may write the final form
\bea
G[{\bf P}, {\bf Q}_k, z] &=& \sum_{k=0}^{\infty} \left(\frac{z}{1+z} \right)^{k+1} 
\sum_{j=0}^{k} (-1)^j \binom{k}{j}  e^{-\frac{j+1}{2} |{\bf Q}_k^T\cdot {\bf P}|^2}.\nonumber \\
\eea
When carrying out integrals involving products of $G$ functions, one may expand such products
in powers of $\frac{z}{1+z}$ using the above expression, such that the resulting integrals are always sums
of Gaussian integrals. We will refer to such an approach to integration as the large-$z$ expansion (LZE),
although it is {\it not} an expansion around $1/z = 0$.
One of the crucial advantages of the LZE over purely numerical integration techniques is that the
latter are usually based on stochastic methods and therefore incur a statistical uncertainty. Furthermore,
such methods are usually confined to integer spatial dimensions. The LZE, on the other hand, is
systematic and can be used to evaluate integrals in arbitrary dimensions, even complex.
We make use of this property below. 
To further display the appealing properties of the LZE, we show in Fig.~\ref{fig:LZEvsVE} its
behavior at second and fourth orders when compared with the virial expansion.
%

%%%%%%%%%%%%%%%%%%%%%%%%%%%%%%%%%%%%%%%%%
\section{Results}
\label{sec:results}

In this section we present our results, focusing on the universal thermodynamics of 
spin-$1/2$ fermions at unitarity. This is a system that is approximately realized in dilute neutron matter
in the crust of neutron stars, and it is realized to great accuracy in ultracold atom experiments carried
out by many groups around the world. Because of the wide interest in this problem from the condensed-matter,
nuclear, and atomic physics communities, this system has been actively studied over the last two decades
with progressively more accuracy, both theoretically as well as experimentally. Our purpose here is
not to provide a more accurate answer for the thermodynamics but rather to use known results as a benchmark
to assess the quality and range of validity of our method.

\subsection{Pressure at unitarity}

\begin{figure}[h]
	\centering
	\includegraphics[width=\linewidth]{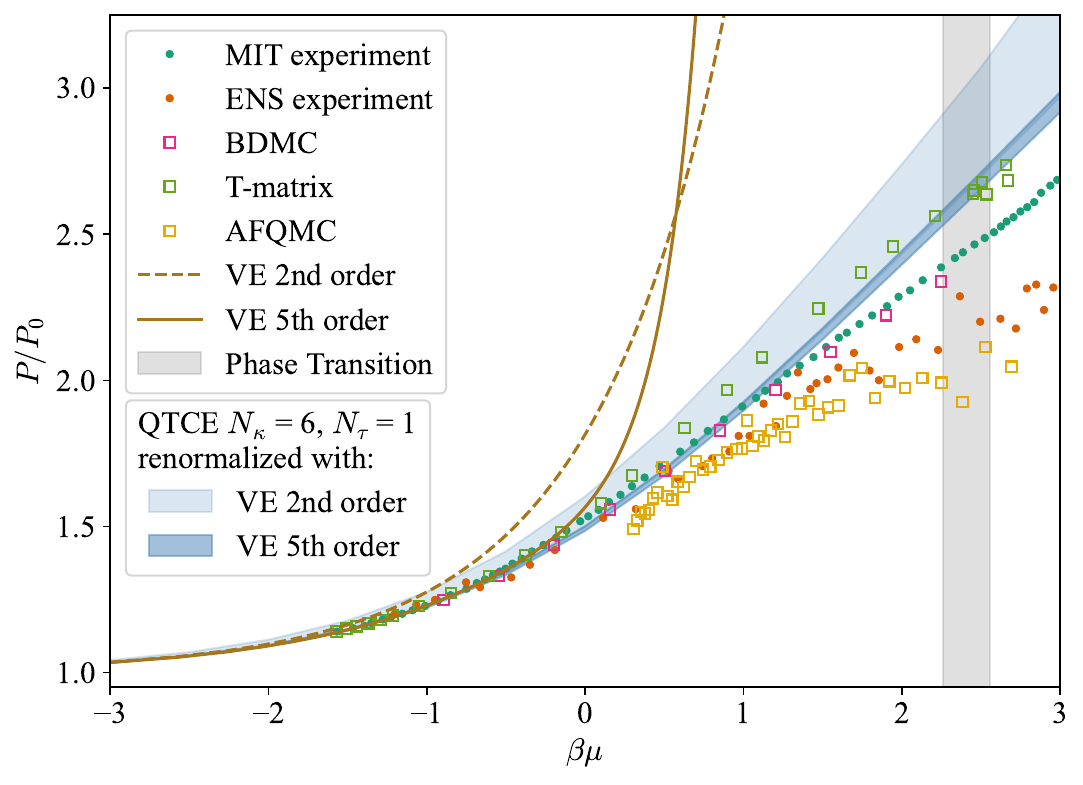}    
	\caption{
		\label{fig:PressureEoS}
		Pressure equation of state compared with other approaches.
		The VE is shown at second order with a dashed gold line and at fifth order~\cite{Hou2020PRL} 
		with a solid gold line. The experimental results from Ref.~\cite{MITExp} are shown with green dots
		and those of Ref.~\cite{ENS-Exp} with red dots. The bold-diagrammatic Monte Carlo (BDMC) results of 
		Ref.~\cite{VanHoucke2012} appear as pink empty squares, the auxiliary-field quantum Monte Carlo (AFQMC)
		data of Ref.~\cite{PhysRevLett.96.090404} is shown as orange empty squares, and the self-consistent results of Ref.~\cite{PhysRevA.75.023610}
		are given by the green empty squares. The vertical gray line around $\beta \mu = 2.4$ shows
		the location of the superfluid phase transition bounded by values corresponding to a critical temperature of 0.152(7) \cite{PhysRevLett.96.160402} and 0.171(5) \cite{PhysRevA.82.053621}. Finally, the QTCE results are shown as light- and 
		dark-shaded blue bands, correspondingly marking the effects of renormalizing with the second-
		and fifth-order VE; the band limits are set by carrying out said renormalization at $\beta \mu = -3.0$ 
		and $\beta \mu = -1.0$ in each case.
		}
\end{figure}

As a first application of the QTCE, we show in Fig.~\ref{fig:PressureEoS} the 
pressure equation of state of spin-$1/2$ fermions at unitarity,
compared with prior data from experiments as well as selected numerical approaches (including self-consistent diagrammatic~\cite{PhysRevA.75.023610} and Monte Carlo methods~\cite{PhysRevLett.96.090404}). Specifically, we show in this plot the variation in our results when renormalizing
using the second-order virial expansion (VE, shown in dashed gold line) or its fifth-order counterpart (solid gold line); 
the corresponding QTCE results are shown as light- and dark-shaded blue bands, respectively. The limits of the QTCE bands are
set by renormalizing at $\beta \mu = -3.0$ and $\beta \mu = -1.0$; we find that the fifth-order VE renormalization
scheme yields a much more constrained band than the second-order scheme. Remarkably, our results match the 
MIT~\cite{MITExp} experimental results up to at least $\beta\mu = 1.0$ (with a slight undershooting deviation around 
$\beta \mu=0$). We emphasize that the QTCE answer is given by a plain (no resummation) partial sum of the cumulant 
expansion of Eq.~(\ref{Eq:Kexpansion}) up to $N^{}_\kappa = 6$ with the most rudimentary approximation for the imaginary-time direction, namely $N^{}_\tau = 1$.
%

%%%%%%%%%%%%%%%
\subsection{Pressure in the BCS-BEC crossover}

Extending our results beyond unitarity, in Fig.~\ref{fig:PressureEoS} we show the 
pressure as a function of $\beta \mu > 0$, across the BCS-BEC crossover, for varying couplings 
parametrized by the ratio $\Delta b_2 / \Delta b_2^\text{UFG}$ (shown with different colors). 
Here, $\Delta b_2$ is the interaction-induced change in the second-order virial coefficient,
which is in one-to-one correspondence with the interaction strength as measured 
usually by the scattering length; $\Delta b_2^\text{UFG} = 1/\sqrt{2}$ is the value in
the unitary limit. In addition, the plot shows partial sums of the cumulant expansion up to 
order $N^{}_\kappa$ (shown with different shades of the same color), indicating the highest power of $\lambda$ 
summed in the series of Eq.~(\ref{Eq:Kexpansion}). All of the results shown in this plot correspond to 
$N^{}_\tau = 1$, i.e. the simplest possible Suzuki-Trotter factorization (i.e. the most coarse temporal lattice). For $\beta\mu < 0$, there is no noticeable difference
from varying $N^{}_\kappa$, which means that the first cumulant, properly renormalized, is enough to capture the behavior of the curve.

The weakest coupling we considered is $\Delta b_2 / \Delta b_2^\text{UFG} = 0.01$,
which is indistinguishable from the noninteracting limit in the scale shown in the plot.
As the coupling is increased, the partial sums display oscillatory behavior at sufficiently
large $\beta\mu$, indicating that the series must be resummed using methods such
as Pade approximants. In particular, we note that the splitting of the partial sums
coincides roughly (and not surprisingly) with the location of the superfluid phase
transition (gray vertical band) around $\beta \mu = 2.4$ in the unitary limit $\Delta b_2 / \Delta b_2^\text{UFG} = 1$
(blue lines).
\begin{figure}[h]
	\centering
	\includegraphics[width=\linewidth]{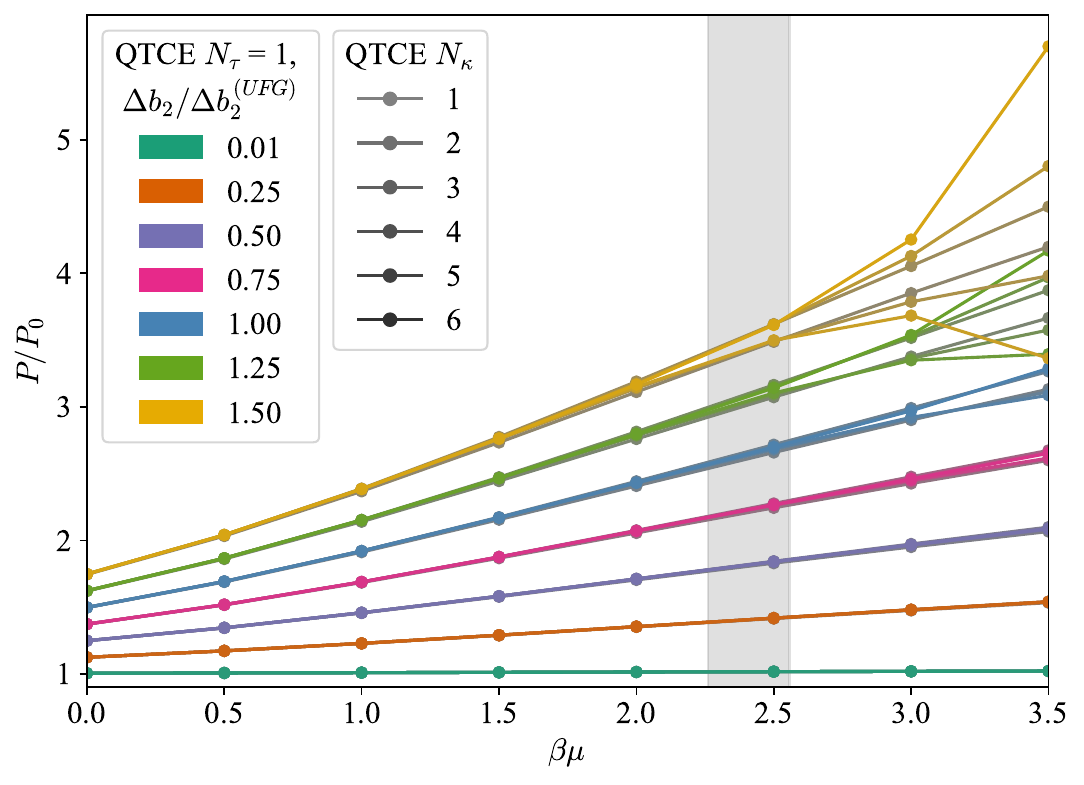}    
	\caption{
		\label{fig:PressureCrossover}
		Pressure $P$, in units of the noninteracting pressure $P_0$, as a function of the coupling 
		$\Delta b_2 / \Delta b_2^\text{UFG}$ (shown in different colors), as obtained using the QTCE
		with $N^{}_\tau = 1$ and varying cumulant expansion order $N^{}_\kappa$ (with the highest color saturation indicating $N^{}_\kappa=6$).
		}
\end{figure}
%

%%%%%%%%%%%%%%%%%%%%%%%%%%%%%
\subsection{Beyond the pressure: density, compressibility and heat capacity at unitarity.}

To characterize the thermodynamics of a system in a more detailed fashion, here we take a closer
look at derivatives of the pressure, which yield the density equation of state and the elementary
response functions given by the compressibility and heat capacity.

The density equation of state is accessed from Eq.~(\ref{Eq:P-lnZ}) using the thermodynamic relation
\beq
N = \pdv{\ln\CZ}{\ln z} = \pdv{\ln\CZ}{(\beta\mu)}.
\eeq

\begin{figure}[h]
	\centering
	\includegraphics[width=\linewidth]{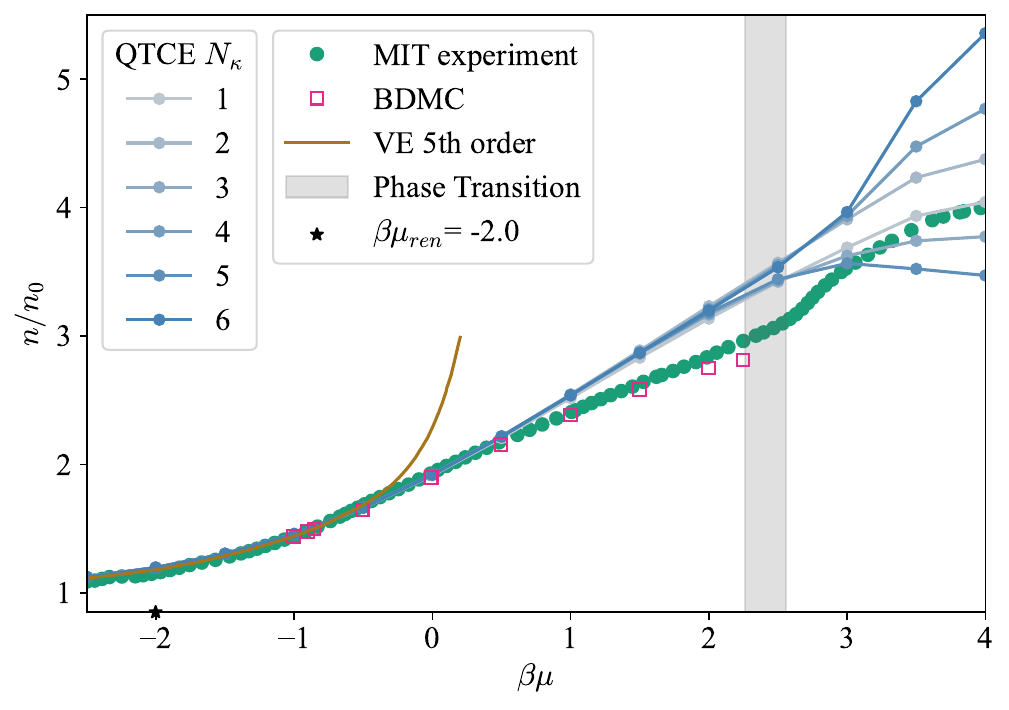}    
	\caption{
		\label{fig:DensityVsBetaMu}
		Density equation of state as a function of the dimension, for various chemical potentials.
		The $N_\tau = 1$ QTCE results are shown in shades of blue according to the maximum cumulant order
		$N_\kappa$ included in the partial sums. The results of the MIT experiment of Ref.~\cite{MITExp}
		are shown with green circles. The diagrammatic Monte Carlo results of Ref.~\cite{VanHoucke2012} are shown with
		pink squares. The virial expansion results at fifth order are shown in gold~\cite{Hou2020PRL}.
		}
\end{figure}

Taking further derivatives of the pressure, one may obtain the isothermal compressibility via the thermodynamic relation
\beq
\kappa_T = -\frac{1}{V}\tpderiv{V}{P}_T = \frac{V\beta}{N^2}\pdv{N}{(\beta\mu)}
\eeq
Our results for the density in the unitary limit are shown in Fig.~\ref{fig:DensityVsBetaMu}.
It is easy to see that, as for the pressure, the QTCE results, even at $N_\tau = 1$, are a dramatic
improvement over the fifth-order VE. As in Fig.~\ref{fig:PressureCrossover}, it is easy to see that
as the superfluid phase transition is approached from low fugacity, the cumulant expansion 
displays wild oscillations that must be resummed.

Finally, the heat capacity per particle at unitarity is given by
\beq
\frac{C_V}{N} = \frac{1}{N}\left(\frac{\partial E}{\partial T}\right)_{N,V} = \frac{3}{5} \frac{d(P/P_{0,T=0})}{d(\beta E_F)^{-1}},
\eeq
where the first equality is the definition and the second equality is valid only at unitarity, where one may use the 
pressure-energy relation $E = 3PV/2$; furthermore, we used $P_{0,T=0} = {2n E_F}/{5}$, where 
$E_F = (3\pi^2 n)^{2/3}/2$ is the Fermi energy of a noninteracting gas at density $n$. 
\begin{figure}[h]
	\centering
	\includegraphics[width=\linewidth]{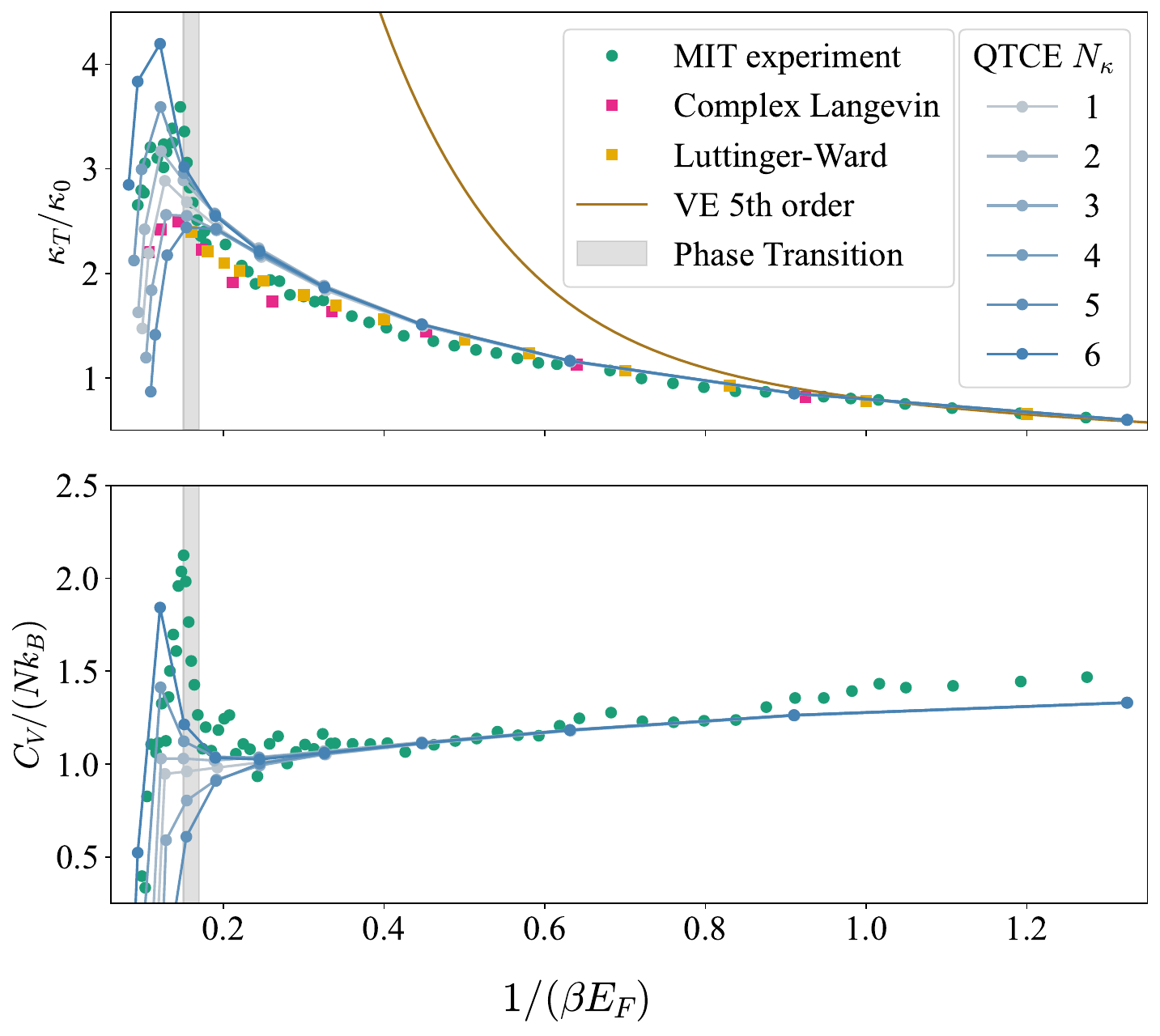}    
	\caption{
		\label{fig:Compressibility}
		{\bf Top}: Compressibility $\kappa_T$, in units of the ground-state noninteracting, zero temperature result 
		$\kappa_0 = 3/(2nE_F)$ as a function of the dimensionless temperature $T/E_F = 1/(\beta E_F)$.
		The green dots show the experimental results of Ref.~\cite{MITExp}, the pink squares show the complex Langevin
		results of Ref.~\cite{PhysRevLett.121.173001}, 
		and the orange squares show the Luttinger-Ward results of Ref.~\cite{PhysRevLett.109.195303}. The solid gold line
		indicates the result of the fifth-order virial expansion, using the results of Ref.~\cite{Hou2020PRL}.
		{\bf Bottom}: Heat capacity as a function of temperature.
		In both plots, the QTCE results are shown with full dots connected by straight lines, with
		shades of blue indicating the maximum order $N^{}_\kappa$ summed in the cumulant expansion.
		The location of the superfluid phase transition is shown as a vertical gray band.
		}
\end{figure}
Our results for the compressibility and heat capacity are shown in Fig.~\ref{fig:Compressibility}.
Both of these response functions show the correct qualitative features for temperatures
above the superfluid transition. Moreover, the agreement with other approaches is arguably
not merely qualitative. However, as with other quantities discussed above, dramatic oscillations
appear in the cumulant expansion in low-temperature phase. This indicates that future analyses will require 
resummation approaches to interpret QTCE data when approaching a phase transition, which is
not surprising.

%%%%%%%%%%%%%%%%%%%%%%%%%%%%%
\subsection{Continuous dimension at unitarity}

\begin{figure}[h]
	\centering
	\includegraphics[width=\linewidth]{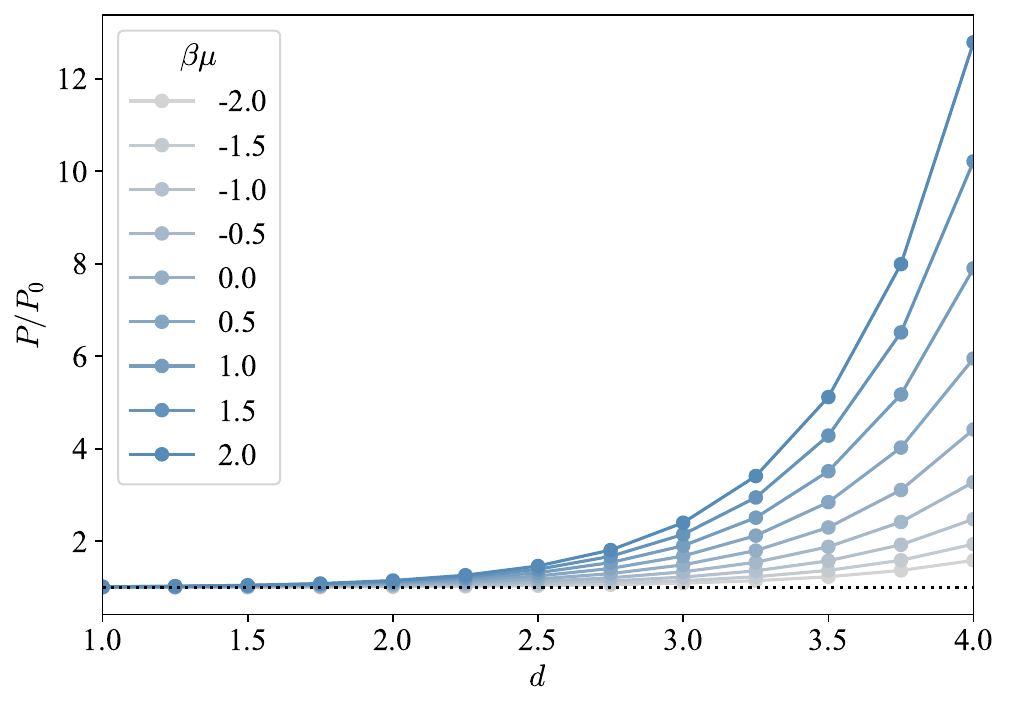}    
	\caption{
		\label{fig:PressureEoSvsdimension}
		Pressure equation of state as a function of the dimension, for various chemical potentials,
		tuned to unitarity in 3D using the fifth-order VE.}
\end{figure}

One of the distinguishing aspects of QTCE is that it encodes the spatial dimension $d$ as an analytic 
variable that can be tuned continuously.
To exemplify that capability, we show in Fig.~\ref{fig:PressureEoSvsdimension} the results obtained for the pressure ratio $P/P_0$ as a function
of $d$, for various chemical potentials, tuned to the unitary limit by renormalizing using the 
condition of matching at low fugacities with the fifth-order VE, as calculated in Ref.~\cite{Hou2020PRL}, across dimensions.
We emphasize that we do not imply that this condition defines a unitary point in different dimensions; rather, it
simply provides a way to formally renormalize the problem as $d$ is varied.

As seen in Fig.~\ref{fig:PressureEoSvsdimension}, the interaction effects on $P$ disappear as $d$ is decreased. Put differently, weaker interactions are more easily able to 
reproduce the condition $\Delta b_2 = 1/\sqrt{2}$ in low dimensions than in high dimensions.
In a real system, the physical picture behind this property is that a two-body bound state appears
in 1D and 2D with vanishingly small attraction, which can dramatically increase $\Delta b_2$
already at weak coupling, as $\Delta b_2$ contains a term that increases {\it exponentially} with the binding energy.

\begin{figure}[h]
	\centering
	\includegraphics[width=\linewidth]{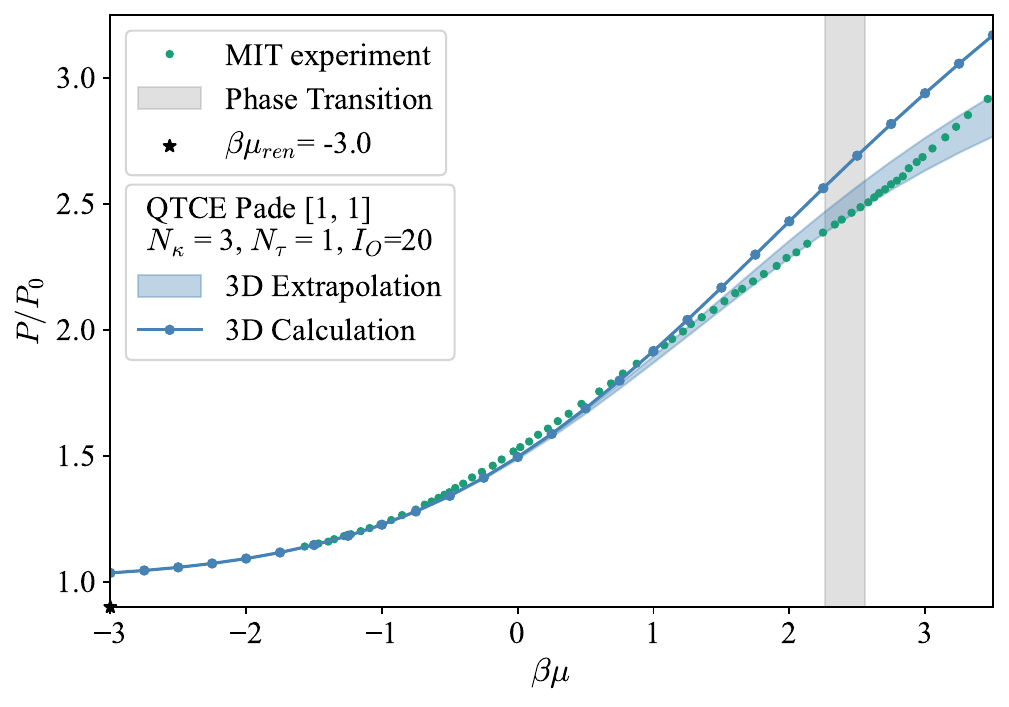}    
	\includegraphics[width=\linewidth]{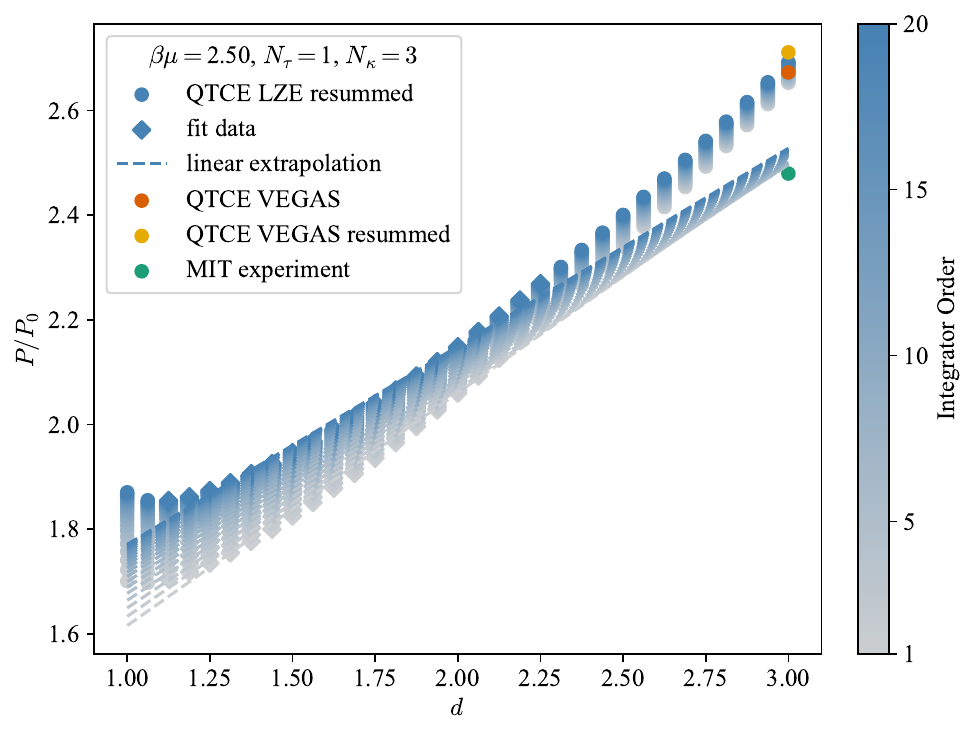}    
	\caption{
		\label{fig:dimExtrapEOS}
		{\bf Top panel:} Pressure equation of state as a function of $\beta\mu$ at unitarity.
		The green dots show the MIT data of Ref.~\cite{MITExp}.
		The gray vertical region shows the superfluid phase transition.
		The star shows the value of $\beta\mu$ used to renormalize with the VE.
		The blue dots joined with a solid line show the results calculated in 3D and resummed using a Pad\'{e} approximant.
		The shaded blue region shows the results calculated by dimensional extrapolation
		between $d=1.125$ and $d=2.0$ (bottom of band) to $d=2.5$ (top of band).
		{\bf Bottom panel:} Spatial-dimension dependence of the pressure equation of state
		at $\beta \mu= 2.5$. The shading from gray to blue indicates the results obtained
		using the LZE integrator at varying orders. 
		}
\end{figure}

The above observation suggests that, when renormalizing using the VE, one may expect the cumulant 
expansion (and more generally the QTCE results in coarse temporal lattices) to present better convergence
properties in low dimensions. To explore that intuition, we show in Fig.~\ref{fig:dimExtrapEOS} (top) the
pressure as a function of $\beta \mu$ at unitarity, obtained with QTCE by dimensional extrapolation using data from 
$d=1.125$ to $d=2.5$, compared with QTCE exactly at $d=3.0$, as well as with the experimental
results from MIT~\cite{MITExp}. The results obtained with QTCE via extrapolation from low dimensions
are clearly in much better quantitative agreement with the experimental numbers than those obtained 
by calculating directly in 3D.

In the bottom panel of Fig.~\ref{fig:dimExtrapEOS}, we provide a more detailed look at the 
dimension dependence of the pressure, at fixed $\beta \mu = 2.5$, for $N_\tau = 1$ and $N_\kappa = 3$.
There is a clear quantitative difference in the final results between taking the QTCE results at face value at $d=3$ 
(regardless of the resummation strategy or integrator used, as shown) and extrapolating from low dimensions (in this case 
between $d=1.125$ and $d=2.25$. 

%

%%%%%%%%%%%%%%%%%%%%%%%%%%%%%
\subsection{Tests in lower dimensions}

\begin{figure}[h]
	\centering
	\includegraphics[width=\linewidth]{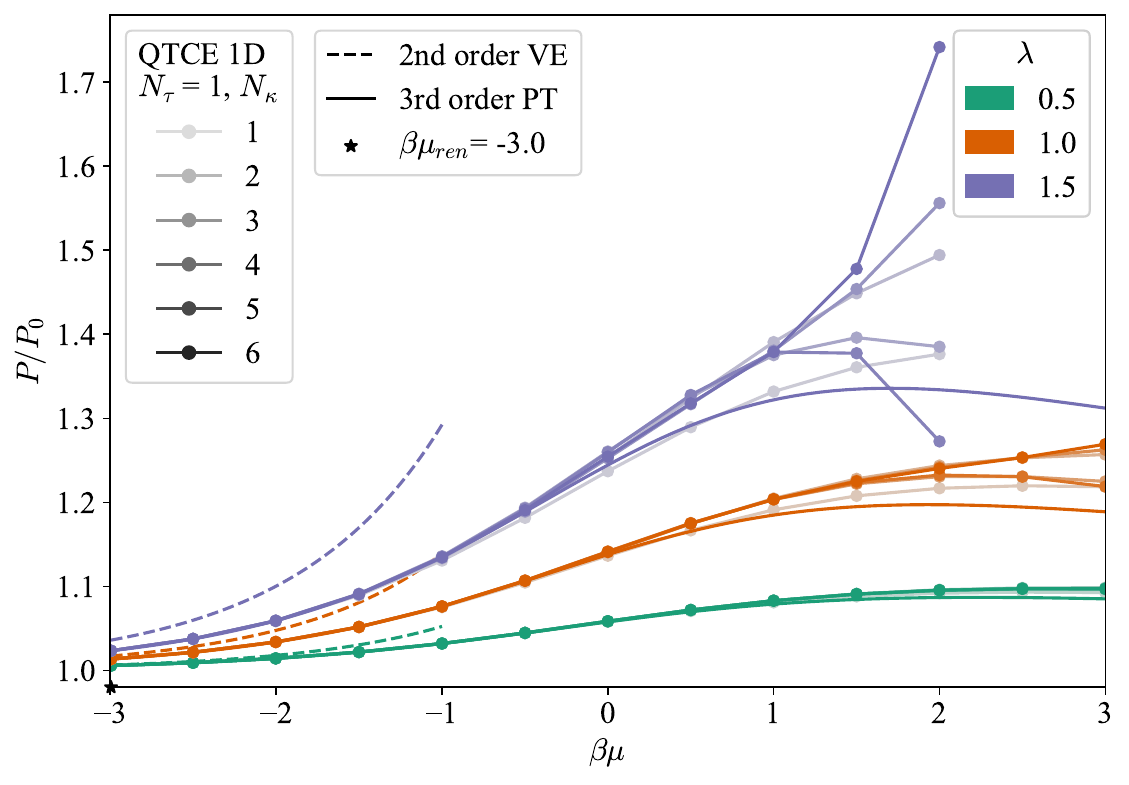}    
	\includegraphics[width=\linewidth]{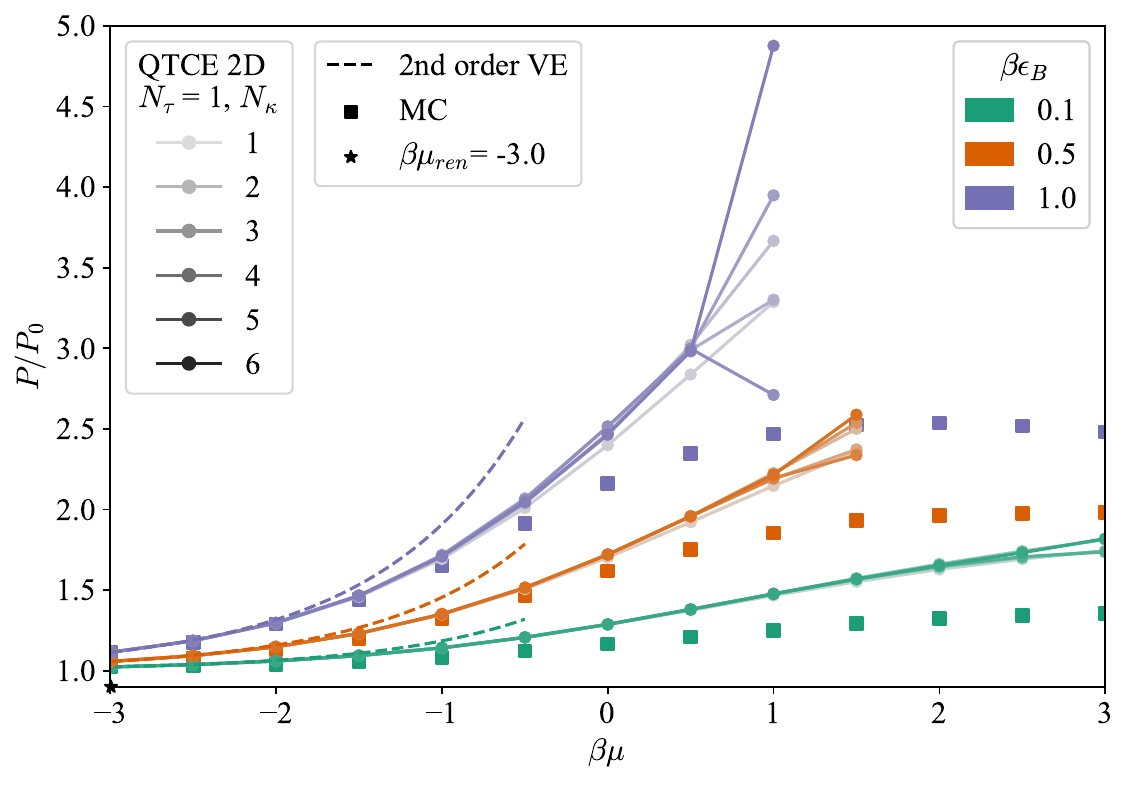}    
	\caption{
		\label{fig:VaryCouplingsLowD}
		{\bf Top panel}: pressure $P$ in units of the noninteracting pressure $P_0$, as a function of $\beta \mu$ for an attractively interacting 1D spin-$1/2$ Fermi gas. The solid lines with data points show the QTCE results; the purely solid lines show third-order perturbation theory, as obtained by Ref.~\cite{PhysRevD.95.094502}; and the dashed lines show the second order virial expansion. The different colors identify different couplings as parametrized by $\lambda$ and the different shades show varying orders in the QTCE cumulant expansion, as parametrized by $N^{}_\kappa$. Finally, the star shows the renormalization point $\beta\mu=-3.0$ used by QTCE to calculate the coupling.
		{\bf Bottom panel}: Same as top panel for the 2D Fermi gas. Instead of perturbation theory results, auxiliary field quantum Monte Carlo calculations are shown, as obtained by Ref.~\cite{PhysRevLett.115.115301}. The coupling is parametrized by $\beta \epsilon^{}_B$, where $\epsilon^{}_B$ is the binding energy of the two-body problem.}
\end{figure}

For completeness, we show in Fig.~\ref{fig:VaryCouplingsLowD} a comparison of QTCE results at $N^{}_\tau=1$ with a third-order
perturbation theory calculation (in 1D; Ref.~\cite{PhysRevD.95.094502}) and auxiliary field quantum Monte Carlo calculations (in 2D; Ref.~\cite{PhysRevLett.115.115301}) for the pressure as a function
of $\beta \mu$. In 1D, the perturbation theory results are used to renormalize the QTCE calculation at $\beta\mu=-3.0$.
In 2D, on the other hand, the second order virial expansion is used to renormalize at $\beta\mu=-3.0$.
Also in both cases, it is clear that the results of QTCE at $N^{}_\tau=1$ are superior to those of the virial expansion
when increasing $\beta\mu$, even in the simplest possible case of $N^{}_\kappa=1$. At sufficiently large $\beta\mu$,
however, the partial sums of the QTCE cumulant expansion break down; in particular, they do so earlier in $\beta\mu$
for stronger couplings, as measured by the parameter $\lambda=2\sqrt{\beta}/a_0$ in 1D (unrelated to the QTCE parameter $\lambda$) 
and $\beta \epsilon^{}_B$ in 2D,  where $a_0$ is the 1D scattering length and $\epsilon^{}_B$ is the binding energy of the two-body system.
%

%%%%%%%%%%%%%%%%%%%%%%%%%%%%%%%%%%%%%%%%%
\section{Beyond $N^{}_\tau=1$ and integrator effects}
\label{sec:NtauEffects}

In the previous sections we have shown results that reflect the successes and shortcomings of 
calculating in coarse temporal lattices. Naturally, it is desirable to remove such discretization effects
altogether, e.g. by increasing $N^{}_\tau$. For each diagram in our calculations, the cost of increasing $N^{}_\tau$
is proportional to $N^{\alpha}_\tau$, where $\alpha$ is the number of loop integrals in the diagram.
Thus, it becomes progressively more computationally expensive to increase $N^{}_\tau$ when increasing $N^{}_\kappa$.
For those reasons, we only present here a discussion of $N^{}_\tau$ effects that is very limited (specifically to
$N^{}_\tau < 10$ ) compared with that of lattice Monte Carlo calculations which regularly operate at $N^{}_\tau \sim 200$.

\begin{figure}[h]
	\centering
	\includegraphics[width=\linewidth]{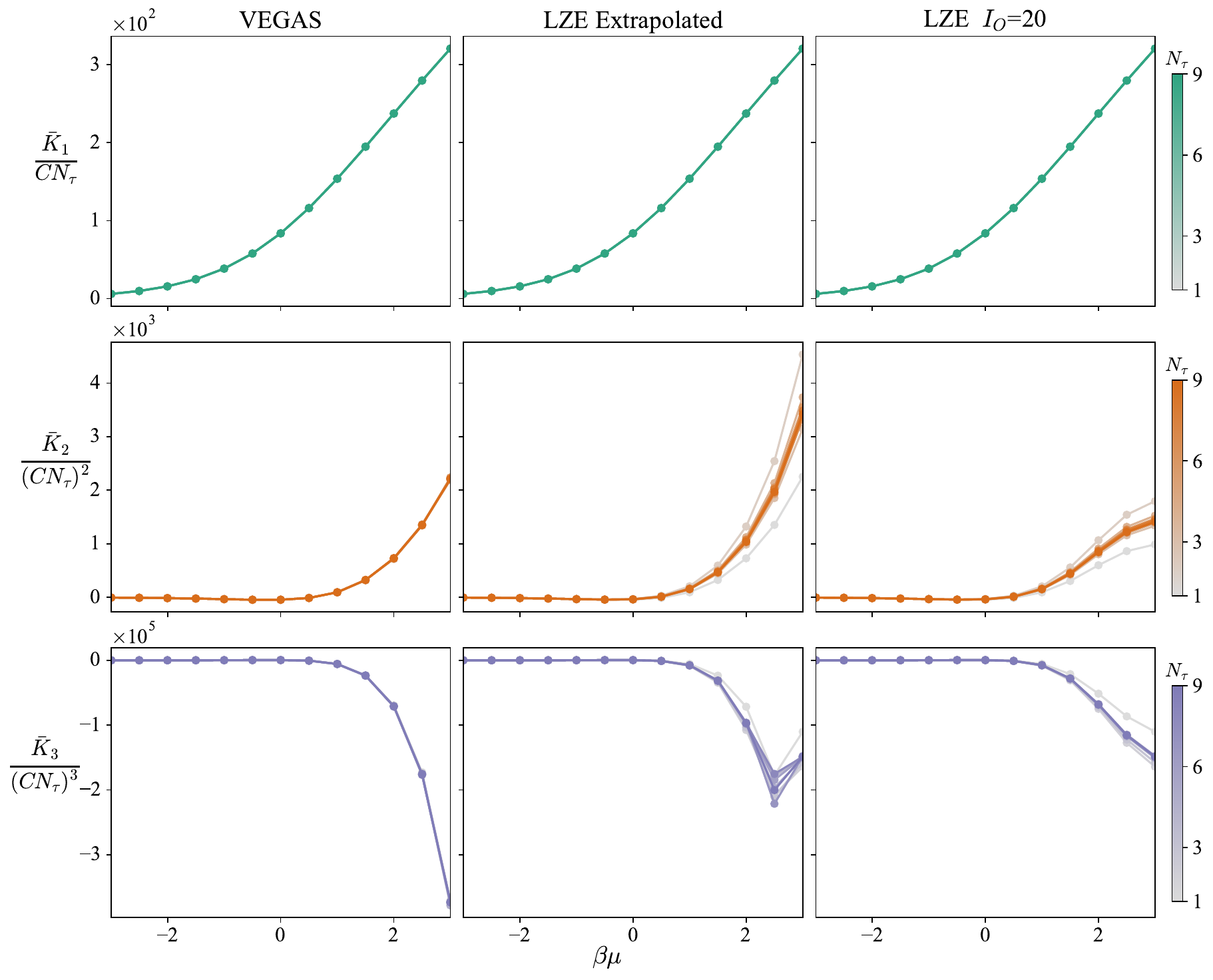}    
	\caption{
		\label{fig:NtauEffects}
		$N^{}_\tau$ effects on the assembled cumulants in $d=3$, as a function of $\beta \mu$, for two
		different integrators: VEGAS (left column) and LZE at order 20 (right column). Note that 
		we include $\bar K^{}_1/N^{}_\tau$ as a reference case that is actually $N^{}_\tau$-independent and calculated exactly
		using Simpson integration methods. Note that $\bar K^{}_k$ is proportional to $C^k$, which 
		has also been factored out in order to obtain the coupling-independent quantities
		shown here.
		}
\end{figure}

In Fig.~\ref{fig:NtauEffects}, we show the effect of increasing $N^{}_\tau$ on the assembled cumulants $\bar K^{}_n$ (adding up
all diagrammatic contributions up to order $N^{}_\kappa = 3$ at $d=3$). Note that the values presented here are coupling-independent: they 
reflect the diagrams' structure induced by the theory and their dependence on $\beta\mu$.

As can be seen in the figure, the first row showing ${\bar K}_1$ displays no variation across integrators,
simply because the required integrals are simple enough to be calculated by QTCE via Gaussian quadratures.
In contrast, there is a clear variation in integrator approach when calculating ${\bar K}_2$ and ${\bar K}_3$:
the stochastic result with VEGAS can differ substantially from the semi-analytical LZE result
(shown at fixed integrator order $I_0 = 20$ in the right column and extrapolated to large integrator order in the
middle column) at large enough $\beta \mu$. In all of these cases, however, the effects due to the 
imaginary-time discretization $N_\tau$ (shown with shades of different colors; see color key on the right side of the plot) are comparatively small: making $N_\tau = 9$ (an extremely modest choice by conventional 
lattice Monte Carlo standards) yields an essentially converged result within a given integrator choice.
Ultimately, the main lesson learned from studying these effects is that, remarkably, finite-temperature 
calculations can produce qualitatively (and in some non-trivial regions also quantitatively) correct 
results already at $N^{}_\tau = 1$.

%%%%%%%%%%%%%%%%%%%%%%%%%%%%%%%%%%%%%%%%%
\section{Conclusion and Outlook}
\label{sec:Conclusion&Outlook}

We have developed a semi-analytic approach to the nonrelativistic quantum many-body problem, which
we call the QTCE. We have explored its capabilities
by comparing with other theoretical approaches as well as experimental results in the unitary
limit of the spin-$1/2$ Fermi gas.

We find that, with a renormalization scheme based on the VE, QTCE can capture the 
thermodynamics of the strongly coupled unitary gas in a remarkably natural way (meaning using mere partial sums 
and no resummation methods) with quantitative agreement with experiments in regions where the virial expansion
fails. Moreover, this is accomplished without driving the number of imaginary time steps $N^{}_\tau$ to a very high number: 
just a single time step captures the correct thermodynamics (pressure, density, compressibility, and heat capacity)
up to $\beta \mu = 1$ (which corresponds to temperatures down to $T/\epsilon^{}_F$ = 0.3. On the other hand, QTCE 
at low $N^{}_\tau$ does not capture the behavior at or beyond the superfluid phase transition ($\beta \mu \simeq 2.4$).
Still, signatures of the transition do appear in QTCE results in the form of dramatic oscillations as the transition is approached 
from high temperatures.

Besides the above, we have probed the behavior of QTCE in 1D and 2D and as a continuous function of dimension (renormalizing 
to the 3D unitary point as a reference), where we also have full analytic control. This feature is interesting as studies of 
dimensional crossover become more common experimentally with ultracold atoms. From the theory standpoint, this is also a 
desirable feature, as the cumulant expansion may display better convergence properties at low or high dimensions, from where 
one may interpolate or extrapolate as needed.

The most straightforward generalizations of QTCE involve including higher spin degrees of freedom (in the form of higher number of species) or other quantum statistics (namely bosons). Beyond those, the generalization to polarized matter (with more than one fugacity variable) is also possible, as is the generalization to interactions beyond the density-density case, although the latter is more computationally intensive than the former. Both, however, would be relevant to tackle the problem of realistic neutron matter, especially if protons are to be included.

%%%%%%%%%%%%%%%%%%%%%%%%%%%%%%%%%%%%%%%%%%%%%%%%
\acknowledgments
We would like to thank Y. Hou for discussions throughout all the stages of this work.
This material is based upon work supported by the
National Science Foundation under Grant No.~PHY{2013078}.

%%%%%%%%%%%%%%%%%%%%%%%%%%%%%%%%%%%%%%%%%%%%%%%%

%\begin{appendix}
%\section{}
%
%
%\end{appendix}

\bibliographystyle{apsrev4-2}
\bibliography{V2Refs}

\begin{thebibliography}{25}%
\makeatletter
\providecommand \@ifxundefined [1]{%
 \@ifx{#1\undefined}
}%
\providecommand \@ifnum [1]{%
 \ifnum #1\expandafter \@firstoftwo
 \else \expandafter \@secondoftwo
 \fi
}%
\providecommand \@ifx [1]{%
 \ifx #1\expandafter \@firstoftwo
 \else \expandafter \@secondoftwo
 \fi
}%
\providecommand \natexlab [1]{#1}%
\providecommand \enquote  [1]{``#1''}%
\providecommand \bibnamefont  [1]{#1}%
\providecommand \bibfnamefont [1]{#1}%
\providecommand \citenamefont [1]{#1}%
\providecommand \href@noop [0]{\@secondoftwo}%
\providecommand \href [0]{\begingroup \@sanitize@url \@href}%
\providecommand \@href[1]{\@@startlink{#1}\@@href}%
\providecommand \@@href[1]{\endgroup#1\@@endlink}%
\providecommand \@sanitize@url [0]{\catcode `\\12\catcode `\$12\catcode
  `\&12\catcode `\#12\catcode `\^12\catcode `\_12\catcode `\%12\relax}%
\providecommand \@@startlink[1]{}%
\providecommand \@@endlink[0]{}%
\providecommand \url  [0]{\begingroup\@sanitize@url \@url }%
\providecommand \@url [1]{\endgroup\@href {#1}{\urlprefix }}%
\providecommand \urlprefix  [0]{URL }%
\providecommand \Eprint [0]{\href }%
\providecommand \doibase [0]{https://doi.org/}%
\providecommand \selectlanguage [0]{\@gobble}%
\providecommand \bibinfo  [0]{\@secondoftwo}%
\providecommand \bibfield  [0]{\@secondoftwo}%
\providecommand \translation [1]{[#1]}%
\providecommand \BibitemOpen [0]{}%
\providecommand \bibitemStop [0]{}%
\providecommand \bibitemNoStop [0]{.\EOS\space}%
\providecommand \EOS [0]{\spacefactor3000\relax}%
\providecommand \BibitemShut  [1]{\csname bibitem#1\endcsname}%
\let\auto@bib@innerbib\@empty
%</preamble>
\bibitem [{\citenamefont {Drut}\ and\ \citenamefont
  {Nicholson}(2013)}]{DrutLattice}%
  \BibitemOpen
  \bibfield  {author} {\bibinfo {author} {\bibfnamefont {J.~E.}\ \bibnamefont
  {Drut}}\ and\ \bibinfo {author} {\bibfnamefont {A.~N.}\ \bibnamefont
  {Nicholson}},\ }\href {https://doi.org/10.1088/0954-3899/40/4/043101}
  {\bibfield  {journal} {\bibinfo  {journal} {J. Phys. G: Nucl. Part. Phys.}\
  }\textbf {\bibinfo {volume} {40}},\ \bibinfo {pages} {043101} (\bibinfo
  {year} {2013})}\BibitemShut {NoStop}%
\bibitem [{\citenamefont {Berger}\ \emph {et~al.}(2021)\citenamefont {Berger},
  \citenamefont {Rammelm\"uller}, \citenamefont {Loheac}, \citenamefont
  {Ehmann}, \citenamefont {Braun},\ and\ \citenamefont
  {Drut}}]{BergerSignProb}%
  \BibitemOpen
  \bibfield  {author} {\bibinfo {author} {\bibfnamefont {C.~E.}\ \bibnamefont
  {Berger}}, \bibinfo {author} {\bibfnamefont {L.}~\bibnamefont
  {Rammelm\"uller}}, \bibinfo {author} {\bibfnamefont {A.~C.}\ \bibnamefont
  {Loheac}}, \bibinfo {author} {\bibfnamefont {F.}~\bibnamefont {Ehmann}},
  \bibinfo {author} {\bibfnamefont {J.}~\bibnamefont {Braun}},\ and\ \bibinfo
  {author} {\bibfnamefont {J.~E.}\ \bibnamefont {Drut}},\ }\href
  {https://doi.org/https://doi.org/10.1016/j.physrep.2020.09.002} {\bibfield
  {journal} {\bibinfo  {journal} {Phys. Rep.}\ }\textbf {\bibinfo {volume}
  {892}},\ \bibinfo {pages} {1} (\bibinfo {year} {2021})}\BibitemShut {NoStop}%
\bibitem [{\citenamefont {Stevenson}(2003)}]{StevensonJMPC}%
  \BibitemOpen
  \bibfield  {author} {\bibinfo {author} {\bibfnamefont {P.~D.}\ \bibnamefont
  {Stevenson}},\ }\href {https://doi.org/10.1142/S0129183103005236} {\bibfield
  {journal} {\bibinfo  {journal} {Int. J. Mod. Phys. C}\ }\textbf {\bibinfo
  {volume} {14}},\ \bibinfo {pages} {1135} (\bibinfo {year}
  {2003})}\BibitemShut {NoStop}%
\bibitem [{\citenamefont {Arthuis}\ \emph {et~al.}(2019)\citenamefont
  {Arthuis}, \citenamefont {Duguet}, \citenamefont {Tichai}, \citenamefont
  {Lasseri},\ and\ \citenamefont {Ebran}}]{ARTHUIS2019202}%
  \BibitemOpen
  \bibfield  {author} {\bibinfo {author} {\bibfnamefont {P.}~\bibnamefont
  {Arthuis}}, \bibinfo {author} {\bibfnamefont {T.}~\bibnamefont {Duguet}},
  \bibinfo {author} {\bibfnamefont {A.}~\bibnamefont {Tichai}}, \bibinfo
  {author} {\bibfnamefont {R.-D.}\ \bibnamefont {Lasseri}},\ and\ \bibinfo
  {author} {\bibfnamefont {J.-P.}\ \bibnamefont {Ebran}},\ }\href
  {https://doi.org/https://doi.org/10.1016/j.cpc.2018.11.023} {\bibfield
  {journal} {\bibinfo  {journal} {Comp. Phys. Comm.}\ }\textbf {\bibinfo
  {volume} {240}},\ \bibinfo {pages} {202} (\bibinfo {year}
  {2019})}\BibitemShut {NoStop}%
\bibitem [{\citenamefont {Arthuis}\ \emph {et~al.}(2021)\citenamefont
  {Arthuis}, \citenamefont {Tichai}, \citenamefont {Ripoche},\ and\
  \citenamefont {Duguet}}]{ARTHUIS2021107677}%
  \BibitemOpen
  \bibfield  {author} {\bibinfo {author} {\bibfnamefont {P.}~\bibnamefont
  {Arthuis}}, \bibinfo {author} {\bibfnamefont {A.}~\bibnamefont {Tichai}},
  \bibinfo {author} {\bibfnamefont {J.}~\bibnamefont {Ripoche}},\ and\ \bibinfo
  {author} {\bibfnamefont {T.}~\bibnamefont {Duguet}},\ }\href
  {https://doi.org/https://doi.org/10.1016/j.cpc.2020.107677} {\bibfield
  {journal} {\bibinfo  {journal} {Comp. Phys. Comm.}\ }\textbf {\bibinfo
  {volume} {261}},\ \bibinfo {pages} {107677} (\bibinfo {year}
  {2021})}\BibitemShut {NoStop}%
\bibitem [{\citenamefont {Tichai}\ \emph {et~al.}(2022)\citenamefont {Tichai},
  \citenamefont {Arthuis}, \citenamefont {Hergert},\ and\ \citenamefont
  {Duguet}}]{Tichai2022}%
  \BibitemOpen
  \bibfield  {author} {\bibinfo {author} {\bibfnamefont {A.}~\bibnamefont
  {Tichai}}, \bibinfo {author} {\bibfnamefont {P.}~\bibnamefont {Arthuis}},
  \bibinfo {author} {\bibfnamefont {H.}~\bibnamefont {Hergert}},\ and\ \bibinfo
  {author} {\bibfnamefont {T.}~\bibnamefont {Duguet}},\ }\href
  {https://doi.org/10.1140/epja/s10050-021-00621-6} {\bibfield  {journal}
  {\bibinfo  {journal} {Eur. Phys. J. A}\ }\textbf {\bibinfo {volume} {58}},\
  \bibinfo {pages} {2} (\bibinfo {year} {2022})}\BibitemShut {NoStop}%
\bibitem [{\citenamefont {Hirata}(2006)}]{Hirata2006}%
  \BibitemOpen
  \bibfield  {author} {\bibinfo {author} {\bibfnamefont {S.}~\bibnamefont
  {Hirata}},\ }\href {https://doi.org/10.1007/s00214-005-0029-5} {\bibfield
  {journal} {\bibinfo  {journal} {Theor. Chem. Acc.}\ }\textbf {\bibinfo
  {volume} {116}},\ \bibinfo {pages} {2} (\bibinfo {year} {2006})}\BibitemShut
  {NoStop}%
\bibitem [{\citenamefont {Hou}\ and\ \citenamefont
  {Drut}(2020{\natexlab{a}})}]{Hou2020PRL}%
  \BibitemOpen
  \bibfield  {author} {\bibinfo {author} {\bibfnamefont {Y.}~\bibnamefont
  {Hou}}\ and\ \bibinfo {author} {\bibfnamefont {J.~E.}\ \bibnamefont {Drut}},\
  }\href {https://doi.org/10.1103/physrevlett.125.050403} {\bibfield  {journal}
  {\bibinfo  {journal} {Phys. Rev. Lett.}\ }\textbf {\bibinfo {volume} {125}},\
  \bibinfo {pages} {050403} (\bibinfo {year} {2020}{\natexlab{a}})}\BibitemShut
  {NoStop}%
\bibitem [{\citenamefont {Hou}\ and\ \citenamefont
  {Drut}(2020{\natexlab{b}})}]{PhysRevA.102.033319}%
  \BibitemOpen
  \bibfield  {author} {\bibinfo {author} {\bibfnamefont {Y.}~\bibnamefont
  {Hou}}\ and\ \bibinfo {author} {\bibfnamefont {J.~E.}\ \bibnamefont {Drut}},\
  }\href {https://doi.org/10.1103/PhysRevA.102.033319} {\bibfield  {journal}
  {\bibinfo  {journal} {Phys. Rev. A}\ }\textbf {\bibinfo {volume} {102}},\
  \bibinfo {pages} {033319} (\bibinfo {year} {2020}{\natexlab{b}})}\BibitemShut
  {NoStop}%
\bibitem [{\citenamefont {Hou}\ \emph {et~al.}(2021)\citenamefont {Hou},
  \citenamefont {Morrell}, \citenamefont {Czejdo},\ and\ \citenamefont
  {Drut}}]{PhysRevResearch.3.033099}%
  \BibitemOpen
  \bibfield  {author} {\bibinfo {author} {\bibfnamefont {Y.}~\bibnamefont
  {Hou}}, \bibinfo {author} {\bibfnamefont {K.~J.}\ \bibnamefont {Morrell}},
  \bibinfo {author} {\bibfnamefont {A.~J.}\ \bibnamefont {Czejdo}},\ and\
  \bibinfo {author} {\bibfnamefont {J.~E.}\ \bibnamefont {Drut}},\ }\href
  {https://doi.org/10.1103/PhysRevResearch.3.033099} {\bibfield  {journal}
  {\bibinfo  {journal} {Phys. Rev. Res.}\ }\textbf {\bibinfo {volume} {3}},\
  \bibinfo {pages} {033099} (\bibinfo {year} {2021})}\BibitemShut {NoStop}%
\bibitem [{\citenamefont {Czejdo}\ \emph {et~al.}(2022)\citenamefont {Czejdo},
  \citenamefont {Drut}, \citenamefont {Hou},\ and\ \citenamefont
  {Morrell}}]{condmat7010013}%
  \BibitemOpen
  \bibfield  {author} {\bibinfo {author} {\bibfnamefont {A.~J.}\ \bibnamefont
  {Czejdo}}, \bibinfo {author} {\bibfnamefont {J.~E.}\ \bibnamefont {Drut}},
  \bibinfo {author} {\bibfnamefont {Y.}~\bibnamefont {Hou}},\ and\ \bibinfo
  {author} {\bibfnamefont {K.~J.}\ \bibnamefont {Morrell}},\ }\href
  {https://doi.org/10.3390/condmat7010013} {\bibfield  {journal} {\bibinfo
  {journal} {Condensed Matter}\ }\textbf {\bibinfo {volume} {7}},\ \bibinfo
  {pages} {13} (\bibinfo {year} {2022})}\BibitemShut {NoStop}%
\bibitem [{\citenamefont {Smith}(1995)}]{RecursiveCumulants}%
  \BibitemOpen
  \bibfield  {author} {\bibinfo {author} {\bibfnamefont {P.~J.}\ \bibnamefont
  {Smith}},\ }\href {https://doi.org/10.1080/00031305.1995.10476146} {\bibfield
   {journal} {\bibinfo  {journal} {The American Statistician}\ }\textbf
  {\bibinfo {volume} {49}},\ \bibinfo {pages} {217} (\bibinfo {year}
  {1995})}\BibitemShut {NoStop}%
\bibitem [{\citenamefont {Negele}\ and\ \citenamefont
  {Orland}(1998)}]{Negele1998Quantum}%
  \BibitemOpen
  \bibfield  {author} {\bibinfo {author} {\bibfnamefont {J.~W.}\ \bibnamefont
  {Negele}}\ and\ \bibinfo {author} {\bibfnamefont {H.}~\bibnamefont
  {Orland}},\ }\href {http://www.worldcat.org/isbn/0738200522} {\emph {\bibinfo
  {title} {Quantum Many-particle Systems}}}\ (\bibinfo  {publisher} {Westview
  Press},\ \bibinfo {year} {1998})\BibitemShut {NoStop}%
\bibitem [{\citenamefont {{Peter Lepage}}(1978)}]{VegasPaper}%
  \BibitemOpen
  \bibfield  {author} {\bibinfo {author} {\bibfnamefont {G.}~\bibnamefont
  {{Peter Lepage}}},\ }\href
  {https://doi.org/https://doi.org/10.1016/0021-9991(78)90004-9} {\bibfield
  {journal} {\bibinfo  {journal} {Journal of Computational Physics}\ }\textbf
  {\bibinfo {volume} {27}},\ \bibinfo {pages} {192} (\bibinfo {year}
  {1978})}\BibitemShut {NoStop}%
\bibitem [{\citenamefont {Haussmann}\ \emph {et~al.}(2007)\citenamefont
  {Haussmann}, \citenamefont {Rantner}, \citenamefont {Cerrito},\ and\
  \citenamefont {Zwerger}}]{PhysRevA.75.023610}%
  \BibitemOpen
  \bibfield  {author} {\bibinfo {author} {\bibfnamefont {R.}~\bibnamefont
  {Haussmann}}, \bibinfo {author} {\bibfnamefont {W.}~\bibnamefont {Rantner}},
  \bibinfo {author} {\bibfnamefont {S.}~\bibnamefont {Cerrito}},\ and\ \bibinfo
  {author} {\bibfnamefont {W.}~\bibnamefont {Zwerger}},\ }\href
  {https://doi.org/10.1103/PhysRevA.75.023610} {\bibfield  {journal} {\bibinfo
  {journal} {Phys. Rev. A}\ }\textbf {\bibinfo {volume} {75}},\ \bibinfo
  {pages} {023610} (\bibinfo {year} {2007})}\BibitemShut {NoStop}%
\bibitem [{\citenamefont {Bulgac}\ \emph {et~al.}(2006)\citenamefont {Bulgac},
  \citenamefont {Drut},\ and\ \citenamefont
  {Magierski}}]{PhysRevLett.96.090404}%
  \BibitemOpen
  \bibfield  {author} {\bibinfo {author} {\bibfnamefont {A.}~\bibnamefont
  {Bulgac}}, \bibinfo {author} {\bibfnamefont {J.~E.}\ \bibnamefont {Drut}},\
  and\ \bibinfo {author} {\bibfnamefont {P.}~\bibnamefont {Magierski}},\ }\href
  {https://doi.org/10.1103/PhysRevLett.96.090404} {\bibfield  {journal}
  {\bibinfo  {journal} {Phys. Rev. Lett.}\ }\textbf {\bibinfo {volume} {96}},\
  \bibinfo {pages} {090404} (\bibinfo {year} {2006})}\BibitemShut {NoStop}%
\bibitem [{\citenamefont {Ku}\ \emph {et~al.}(2012)\citenamefont {Ku},
  \citenamefont {Sommer}, \citenamefont {Cheuk},\ and\ \citenamefont
  {Zwierlein}}]{MITExp}%
  \BibitemOpen
  \bibfield  {author} {\bibinfo {author} {\bibfnamefont {M.~J.~H.}\
  \bibnamefont {Ku}}, \bibinfo {author} {\bibfnamefont {A.~T.}\ \bibnamefont
  {Sommer}}, \bibinfo {author} {\bibfnamefont {L.~W.}\ \bibnamefont {Cheuk}},\
  and\ \bibinfo {author} {\bibfnamefont {M.~W.}\ \bibnamefont {Zwierlein}},\
  }\href {https://doi.org/10.1126/science.1214987} {\bibfield  {journal}
  {\bibinfo  {journal} {Science}\ }\textbf {\bibinfo {volume} {335}},\ \bibinfo
  {pages} {563} (\bibinfo {year} {2012})},\ \Eprint
  {https://arxiv.org/abs/https://www.science.org/doi/pdf/10.1126/science.1214987}
  {https://www.science.org/doi/pdf/10.1126/science.1214987} \BibitemShut
  {NoStop}%
\bibitem [{\citenamefont {Nascimb\`{e}ne}\ \emph {et~al.}(2010)\citenamefont
  {Nascimb\`{e}ne}, \citenamefont {Navon}, \citenamefont {Jiang}, \citenamefont
  {Chevy},\ and\ \citenamefont {Salomon}}]{ENS-Exp}%
  \BibitemOpen
  \bibfield  {author} {\bibinfo {author} {\bibfnamefont {S.}~\bibnamefont
  {Nascimb\`{e}ne}}, \bibinfo {author} {\bibfnamefont {N.}~\bibnamefont
  {Navon}}, \bibinfo {author} {\bibfnamefont {K.~J.}\ \bibnamefont {Jiang}},
  \bibinfo {author} {\bibfnamefont {F.}~\bibnamefont {Chevy}},\ and\ \bibinfo
  {author} {\bibfnamefont {C.}~\bibnamefont {Salomon}},\ }\href
  {https://doi.org/10.1038/nature08814} {\bibfield  {journal} {\bibinfo
  {journal} {Nature}\ }\textbf {\bibinfo {volume} {463}},\ \bibinfo {pages}
  {1057} (\bibinfo {year} {2010})}\BibitemShut {NoStop}%
\bibitem [{\citenamefont {Houcke}\ \emph {et~al.}(2012)\citenamefont {Houcke},
  \citenamefont {Werner}, \citenamefont {Kozik}, \citenamefont {Prokof'ev},
  \citenamefont {Svistunov}, \citenamefont {Ku}, \citenamefont {Sommer},
  \citenamefont {Cheuk}, \citenamefont {Schirotzek},\ and\ \citenamefont
  {Zwierlein}}]{VanHoucke2012}%
  \BibitemOpen
  \bibfield  {author} {\bibinfo {author} {\bibfnamefont {K.~V.}\ \bibnamefont
  {Houcke}}, \bibinfo {author} {\bibfnamefont {F.}~\bibnamefont {Werner}},
  \bibinfo {author} {\bibfnamefont {E.}~\bibnamefont {Kozik}}, \bibinfo
  {author} {\bibfnamefont {N.}~\bibnamefont {Prokof'ev}}, \bibinfo {author}
  {\bibfnamefont {B.}~\bibnamefont {Svistunov}}, \bibinfo {author}
  {\bibfnamefont {M.~J.~H.}\ \bibnamefont {Ku}}, \bibinfo {author}
  {\bibfnamefont {A.~T.}\ \bibnamefont {Sommer}}, \bibinfo {author}
  {\bibfnamefont {L.~W.}\ \bibnamefont {Cheuk}}, \bibinfo {author}
  {\bibfnamefont {A.}~\bibnamefont {Schirotzek}},\ and\ \bibinfo {author}
  {\bibfnamefont {M.~W.}\ \bibnamefont {Zwierlein}},\ }\href
  {https://doi.org/10.1038/nphys2273} {\bibfield  {journal} {\bibinfo
  {journal} {Nature Physics}\ }\textbf {\bibinfo {volume} {8}},\ \bibinfo
  {pages} {366} (\bibinfo {year} {2012})}\BibitemShut {NoStop}%
\bibitem [{\citenamefont {Burovski}\ \emph {et~al.}(2006)\citenamefont
  {Burovski}, \citenamefont {Prokof'ev}, \citenamefont {Svistunov},\ and\
  \citenamefont {Troyer}}]{PhysRevLett.96.160402}%
  \BibitemOpen
  \bibfield  {author} {\bibinfo {author} {\bibfnamefont {E.}~\bibnamefont
  {Burovski}}, \bibinfo {author} {\bibfnamefont {N.}~\bibnamefont {Prokof'ev}},
  \bibinfo {author} {\bibfnamefont {B.}~\bibnamefont {Svistunov}},\ and\
  \bibinfo {author} {\bibfnamefont {M.}~\bibnamefont {Troyer}},\ }\href
  {https://doi.org/10.1103/PhysRevLett.96.160402} {\bibfield  {journal}
  {\bibinfo  {journal} {Phys. Rev. Lett.}\ }\textbf {\bibinfo {volume} {96}},\
  \bibinfo {pages} {160402} (\bibinfo {year} {2006})}\BibitemShut {NoStop}%
\bibitem [{\citenamefont {Goulko}\ and\ \citenamefont
  {Wingate}(2010)}]{PhysRevA.82.053621}%
  \BibitemOpen
  \bibfield  {author} {\bibinfo {author} {\bibfnamefont {O.}~\bibnamefont
  {Goulko}}\ and\ \bibinfo {author} {\bibfnamefont {M.}~\bibnamefont
  {Wingate}},\ }\href {https://doi.org/10.1103/PhysRevA.82.053621} {\bibfield
  {journal} {\bibinfo  {journal} {Phys. Rev. A}\ }\textbf {\bibinfo {volume}
  {82}},\ \bibinfo {pages} {053621} (\bibinfo {year} {2010})}\BibitemShut
  {NoStop}%
\bibitem [{\citenamefont {Rammelm\"uller}\ \emph {et~al.}(2018)\citenamefont
  {Rammelm\"uller}, \citenamefont {Loheac}, \citenamefont {Drut},\ and\
  \citenamefont {Braun}}]{PhysRevLett.121.173001}%
  \BibitemOpen
  \bibfield  {author} {\bibinfo {author} {\bibfnamefont {L.}~\bibnamefont
  {Rammelm\"uller}}, \bibinfo {author} {\bibfnamefont {A.~C.}\ \bibnamefont
  {Loheac}}, \bibinfo {author} {\bibfnamefont {J.~E.}\ \bibnamefont {Drut}},\
  and\ \bibinfo {author} {\bibfnamefont {J.}~\bibnamefont {Braun}},\ }\href
  {https://doi.org/10.1103/PhysRevLett.121.173001} {\bibfield  {journal}
  {\bibinfo  {journal} {Phys. Rev. Lett.}\ }\textbf {\bibinfo {volume} {121}},\
  \bibinfo {pages} {173001} (\bibinfo {year} {2018})}\BibitemShut {NoStop}%
\bibitem [{\citenamefont {Enss}\ and\ \citenamefont
  {Haussmann}(2012)}]{PhysRevLett.109.195303}%
  \BibitemOpen
  \bibfield  {author} {\bibinfo {author} {\bibfnamefont {T.}~\bibnamefont
  {Enss}}\ and\ \bibinfo {author} {\bibfnamefont {R.}~\bibnamefont
  {Haussmann}},\ }\href {https://doi.org/10.1103/PhysRevLett.109.195303}
  {\bibfield  {journal} {\bibinfo  {journal} {Phys. Rev. Lett.}\ }\textbf
  {\bibinfo {volume} {109}},\ \bibinfo {pages} {195303} (\bibinfo {year}
  {2012})}\BibitemShut {NoStop}%
\bibitem [{\citenamefont {Loheac}\ and\ \citenamefont
  {Drut}(2017)}]{PhysRevD.95.094502}%
  \BibitemOpen
  \bibfield  {author} {\bibinfo {author} {\bibfnamefont {A.~C.}\ \bibnamefont
  {Loheac}}\ and\ \bibinfo {author} {\bibfnamefont {J.~E.}\ \bibnamefont
  {Drut}},\ }\href {https://doi.org/10.1103/PhysRevD.95.094502} {\bibfield
  {journal} {\bibinfo  {journal} {Phys. Rev. D}\ }\textbf {\bibinfo {volume}
  {95}},\ \bibinfo {pages} {094502} (\bibinfo {year} {2017})}\BibitemShut
  {NoStop}%
\bibitem [{\citenamefont {Anderson}\ and\ \citenamefont
  {Drut}(2015)}]{PhysRevLett.115.115301}%
  \BibitemOpen
  \bibfield  {author} {\bibinfo {author} {\bibfnamefont {E.~R.}\ \bibnamefont
  {Anderson}}\ and\ \bibinfo {author} {\bibfnamefont {J.~E.}\ \bibnamefont
  {Drut}},\ }\href {https://doi.org/10.1103/PhysRevLett.115.115301} {\bibfield
  {journal} {\bibinfo  {journal} {Phys. Rev. Lett.}\ }\textbf {\bibinfo
  {volume} {115}},\ \bibinfo {pages} {115301} (\bibinfo {year}
  {2015})}\BibitemShut {NoStop}%
\end{thebibliography}%

\end{document}